\documentclass[amsmath,amssymb,floatfix,superscriptaddress]{revtex4} 
\usepackage[usenames]{xcolor}
\usepackage{graphicx}
\usepackage{dcolumn}  % Align table columns on decimal point
\usepackage{bm}  % bold math
\usepackage{tikz}

\newcommand{\ind}[1]{_\textrm{#1}}

\begin{document}

\title{The one dimensional Coulomb lattice 
fluid capacitor}

\author{Vincent D\' emery}
\affiliation{Laboratoire de Physique Th\' eorique (IRSAMC), Universit\' e de Toulouse, UPS and CNRS, F-31062 Toulouse, France}

\author{David S. Dean}
\affiliation{Universit\'e de  Bordeaux and CNRS, Laboratoire Ondes et Mati\`ere d'Aquitaine (LOMA), UMR 5798, F-33400 Talence, France}

\author{Thomas C. Hammant}
\affiliation{DAMTP CMS, Wilberforce Road, Cambridge CB3 0WA, United Kingdom}

\author{Ronald R. Horgan}
\affiliation{DAMTP CMS, Wilberforce Road, Cambridge CB3 0WA, United Kingdom}

\author{Rudolf Podgornik}
\affiliation{Department of Theoretical Physics, J. Stefan Institute and Department of Physics, Faculty of Mathematics and Physics, University of Ljubljana, SI-1000 Ljubljana, Slovenia}

\begin{abstract}
The one dimensional Coulomb lattice fluid in a capacitor configuration is studied. The model is 
formally exactly soluble via a transfer operator method within a field theoretic representation 
of the model. The only interactions present in the model are the one dimensional Coulomb interaction between cations and anions and the steric interaction imposed by restricting the maximal occupancy at
any lattice site to one particle. Despite the simplicity of the model, a wide range of intriguing physical 
phenomena arise, some of which are strongly reminiscent of those seen in experiments and numerical 
simulations of three dimensional ionic liquid based capacitors. Notably we find regimes where
over-screening and density oscillations are seen near the capacitor plates. The capacitance is also 
shown to exhibit strong oscillations as a function of applied voltage. It is also shown that the corresponding mean field theory misses most of these effects. The analytical results are confirmed by extensive numerical simulations.
\end{abstract}

\maketitle

\section{Introduction}

The physics of interacting Coulomb systems has been extensively studied for decades \cite{poded,pooned,review}. Most of this study has been concentrated on the statistical mechanics of electrolytes such as salt solutions. While there still exists a number of outstanding questions concerning the bulk thermodynamics of  electrolytes, there has been just as intense an effort to understand how the presence of electrolytes modify the interaction between charged surfaces, colloids and macromolecules. The
underlying statistical mechanics problem being to date unsolved, the study of these systems relies on
the use of approximations. The simplest models of electrolytes are called primitive models, where the 
ions are taken to be point-like charges and the only interactions are electrostatic. 

For these simple electrolyte models with charged interfaces, two regimes are
now reasonably well understood and their results have been validated numerically. The first is the so called {\sl weak coupling regime} where the mean field or Poisson-Boltzmann (PB) theory \cite{bo2005} correctly describes the system, ignoring fluctuations about the mean field. These can be taken into account to refine the accuracy of the results. The PB theory
is valid when the distance over which the ions interact with each other (the Bjerrum length) is smaller than the distance at which they interact with the charged surface (the Gouy-Chapman length).
From another mathematical point of view the field theory describing interacting charged particles can be analyzed in the saddle point approximation \cite{RP-SP}, and in the regime of its validity the fluctuations renormalize the mean field results improving its accuracy. The other limiting regime is the so called {\sl strong coupling regime} where the problem can be treated within an effectively virial-like expansion based upon the interaction of  single particles with the charged surface \cite{mo2005}. This regime occurs when the Bjerrum length is much large than the Gouy-Chapman length, which becomes the case as the valency of the ions increases. The underlying instability in a point-like charge model due to the pairing of 
anions and cations means that  the strong coupling approximation has only been applied to the
counter-ion only case in the presence of charged surfaces. The region between these two limiting regimes is less well understood and a theory extrapolating between the two is still lacking. 

Results of calculations in these two limiting regimes do not have to be made finite by invoking a finite size of the ions.  This is because (i) point charges are smeared out by the PB approximation and (ii) as stated above, the strong coupling approximation pertaining to the counter-ion only case, is effectively a one particle theory. However, it is clear that in certain circumstances steric effects due to the finite size of 
ions will come into play, not only especially near highly charged bounding surfaces but also in the bulk for
systems composed of large charged molecules such as room temperature ionic liquids (RTILs). The presence of a new length scale given by the molecular size now complicates the above discussion of
strong and weak coupling regimes, rendering the study of RTILs much more complex than  simple electrolyte models.

RTILs are  fluids with large, asymmetric ions where steric effects cannot be neglected.  These liquids
have a number of  practical technological applications \cite{review1}; for instance, they are used as electrolytes in  fuel and solar cells as well as super-capacitors. The study of bulk properties of 
ionic liquids was stimulated by attempts to understand critical behavior of electrolytes. Here the effect of
finite size ions needs to be taken into account in order to regularize the model. The restricted primitive model incorporates hard core effects in a continuum context and can be studied via approximate methods from liquid theory, for instance integral equations as well as field theoretic approaches \cite{ste95,yeh96, jia02}. Another way to take into account the finite size of ions is to restrict ions to points on a regular lattice (for a historical sketch of lattice Coulomb fluids (LCF) see \cite{Bazant-rev,Andelman-rev}) and forbid that there be more than one particle at any given site
\cite{cia02,kob202,bro02,kob02,art03}. This is clearly a rather idealized model since it assumes an underlying lattice structure which will influence the thermodynamics of the system and also since it treats the effective molecular sizes of cations and anions as the same.

The study of LCFs in the capacitor configuration is a more recent subject of study. It is physically clear  \cite{Kornyshev1} that  the size of the ionic species will restrict the maximal local charge density that can be locally attained in a RTIL, for instance the counter-ion charge density near a highly charged surface will have a maximal  value at closest packing.  In certain regimes therefore, we should expect a fundamentally different behavior of  ionic liquids at charged interfaces compared with  aqueous electrolytes. Not withstanding  this proviso,  aspects of the behavior of  RTIL capacitors  can be captured by a  mean-field treatment of the lattice Coulomb fluid model \cite{Bazant-rev,Kornyshev1}.  Qualitatively the mean-field 
treatment of the LCF predicts that  steric constraints  cause the capacitance of ionic liquids to decay at large applied voltages (owing to the saturation of counter-ions at the capacitor plates). At the point of zero charge (PZC) the capacitance can exhibit a maximum and then decay monotonically, however in certain cases  it can also be a local minimum, increasing to a maximum value at finite charge and then decaying (and is thus a non-monotonic  function of applied voltage). Which of these two capacitance behaviors is predicted by the mean field theory depends on the lattice packing fraction \cite{Kornyshev2}. 
For dilute electrolytes the capacitance at PZC is always a minimum \cite{Kornyshev1}, steric effects which reduce the capacitance only becoming important at higher voltages. The basic LCF model
also uses a rather simple description of the capacitor plates, neglecting dispersion interactions that will come into play for ions near metallic electrodes and will influence the behavior of ions even at 
zero applied voltage \cite{ski10,lot10}. Taking into account these dispersion effects, that essentially lead to binding of discrete ions to their image charges in the metallic plate, leads to significantly different predictions for ionic distributions at the capacitor plates and consequently to significantly enhanced capacitances. The effects of these interactions improve the quantitative agreement between theory and experiment.

At an experimental level, X-ray reflectivity \cite{Mezger}, SFA \cite{SFA} and AFM \cite{AFM} studies at charged interfaces show an alternating charge distribution starting with an {\sl over-screening} cationic layer, at the negatively charged substrate, which decays roughly exponentially into the bulk liquid with a periodicity comparable with the size of ionic species \cite{Perkin}. Similar results are found in numerical simulations \cite{Kornyshev2}.

The  charge  layering seen in experiments and simulations is expected to be a generic feature of RTILs at sufficiently strongly charged interfaces resulting from an interplay between steric effects and strong electrostatic coupling. While the phenomenology of the  differential capacity is described by  the mean-field  solution of the LCF model \cite{Kornyshev2}, the over-screening and alternate charge layering at a capacitor plate are not.  

In order to explain overcharging and charge oscillations, but staying within the mean field approximation, Bazant {\sl et al.}  \cite{Bazant}  proposed a phenomenological theory based on a Landau-Ginzburg-like functional containing the standard LCF free energy \cite{Borukhov} but with an additional higher order potential-gradient term, similar to what is found in Cahn-Hilliard models. The origin of this  higher order gradient term can be justified  \cite{Santangelo} via a decomposition of the Coulomb interaction into a long-distance mean-field-like component and a non-mean-field strong coupling component \cite{Kanduc}. This alternative interpretation of the origin of this  phenomenological term as a result of strong coupling or correlation effects leads to the natural question as to whether it is the model or the approximation that needs to be modified to understand the physics of these systems. The latter possibility suggests that  a more detailed non-mean-field analysis of the original Kornyshev LCF model is appropriate. Therefore, in this paper, we will study the exact solution of
the one dimensional (1D) Coulomb fluid or gas where all the physics of the basic LCF model can be analyzed. While the physics of this system will be different to the 3D system, it is of interest because
we can assess the validity of mean field theory in this case by comparing it with an exact solution. In addition the 1D LCF corresponds to sheets of charge (at discrete lattice 
sites), and as experimental results and simulations suggest that a layering phenomena is present, one might hope that this model does contain and possibly explain some of the phenomena of the original 3D system. Indeed we shall see that the phenomena of over-screening, charge oscillations and some of the more exotic behavior of the capacitance of these systems are predicted by our exact solution. The conclusion of the paper is that it is quite possible that some of the physics of RTIL capacitors can be explained by the LCF model if the model is properly treated at the mathematical level.

We thus propose an exact  analysis, albeit in 1D, that demonstrates the full physical phenomenology  of the LCF model, and in particular show that charge density oscillations, over-screening, and oscillations in the capacitance  emerge naturally from this model without the introduction of any new physical interactions. A short version of this work can be found in \cite{v1}, the version here fully details the calculations as well as presents a new mathematical analysis of the exact solution, in particular to explain the surface charging through a fixed applied voltage for large capacitor plates. The physical properties of the systems are discussed in greater detail as well as the relationship between the exact result and the corresponding mean-field theory. Finally, given the rather rich and exotic behavior found in our analytical approach we have confirmed our analytic results 
by comparing them with extensive Monte Carlo simulations of the 1D LCF. 

\section{One dimensional lattice Coulomb fluid model}

Here we describe the one dimensional LCF and show how it can be formally solved exactly. The first exact solutions of the one dimensional Coulomb fluid or gas are due to Lenard and Edwards
\cite{edlen} where the bulk properties of a point-like model of ions was studied. The problem was then  analyzed in a more mathematical context \cite{aiz}, and the peculiar dependence of the bulk free energy 
on the electric field applied to such a system was elucidated. Later the effective interactions between
charged surfaces in this model was studied \cite{dd1d,dd1d2} in the presence of counter and co-ions and
counter-ions only. These studies were also used to benchmark the range of validity of the various
approximation schemes (strong and weak coupling) discussed in the Introduction. Here we present a formally exact solution to the LCF in one dimension based on a functional transfer matrix formalism which can be used to extract, numerically, all thermodynamical observables to any required numerical
precision.    

\subsection{Without external charges}

We consider a one dimensional system where  sheets of density $\sigma_*$ (cations) or $-\sigma_*$ (anions)
can sit at points on a lattice of size $M$ and lattice spacing $a$.  If we assume that the system has an effective surface area $A$, the effective one dimensional Coulomb Hamiltonian can be written as
\begin{equation}
H = -{A\sigma_*^2 a\over 4\epsilon} \sum_{i,j=0}^{M-1} |i-j|S_i S_j
\end{equation}
where $\epsilon$ is the dielectric constant within the bulk of the ionic liquid.

\begin{figure}
\begin{center}
\includegraphics[width=.9\linewidth]{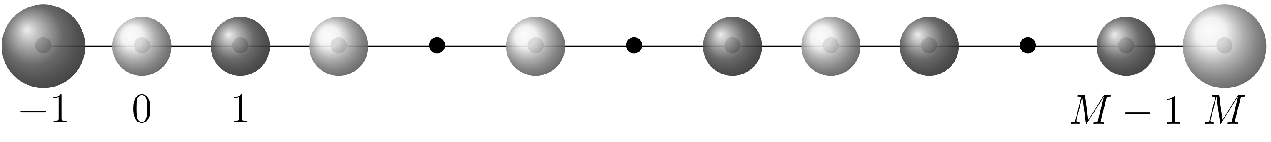}
\end{center}
\caption{Lattice Coulomb fluid model. Dark balls stand for cations and light balls for anions. Boundary charges are at sites $-1$ and $M$.}
\label{scheme_lcg}
\end{figure}

In the above,  $S_i$ is a classical spin taking the value $S_i=1$ if there is a cation at lattice site $i$,
$S_i=-1$ if there is an anion and $S_i=0$ if it is empty. The basic 1D model is shown in 
Fig. \ref{scheme_lcg}. We will impose overall electroneutrality on the 
system which means that $\sum_i S_i=0$ using the discrete Fourier representation
\begin{equation}
\delta_{\sum_i S_i,0} ={1\over 2\pi}\int_{-\pi}^{\pi} d\psi \exp\left(i\psi\sum_{i=0}^{M-1} S_i\right).
\end{equation}
We will work in the grand canonical ensemble where the  grand partition function can be written as 
\begin{equation}
\Xi = {\rm Tr}_{S_i} \left[ \mu^{\sum_{i=0}^{M-1} |S_i|}\int_{-\pi}^{\pi}\exp\left(\frac{\gamma}{2} \sum_{i,j=0}^{M-1} |i-j| S_i S_j + i\psi\sum_{i=0}^{M-1} S_i\right) {d\psi\over 2\pi}\right],
\end{equation}
where $\mu$ is the fugacity of both types of ions (as the liquid is symmetric  between anions and cations). We have also introduced the dimensionless parameter
\begin{equation}\label{length}
\gamma={A\beta \sigma_*^2 a\over 2 \epsilon},
\end{equation}
which is the ratio of the electrostatic energy between two ions at neighboring sites to the typical thermal energy $k_BT = 1/\beta$. Note that the overall charge of an ion in this notation is $q= A \sigma_*$ and in this notation
\begin{equation}
\gamma = {q^2 \beta a\over 2 A \epsilon}.
\end{equation}
Above the trace is over the spin values $S_i$ at each site and the integral over the field $\psi$ enforces the electroneutrality constraint. We notice that the integral over $\psi$ may be translated, it just has to run over an interval of length $2\pi$.
Carrying out a Hubbard-Stratonovitch transformation gives
\begin{equation}
\exp\left(\frac{\gamma}{2} \sum_{i,j=0}^{M-1} |i-j| S_i S_j\right)=N\int[d\phi]\exp\left(-\frac{1}{2\gamma}\int\left[{d\phi(x)\over dx}\right]^2 dx+i\sum_iS_j\phi(j)\right),\label{pf1}
\end{equation}
where $N$ is a normalization constant which can be determined at the end of the computation imposing that the grand partition function without ions ($\mu=0$) should be equal to one. The field $\phi$ is related to the electrostatic potential of the problem. The energy of a point charge $q_i$ in an electrostatic potential $V(x)$ is simply given by ${\cal E}_{ES}= qV(x)$ which for this system gives a factor in the 
exponential Botlzmann weight $-\beta {\cal E}_{ES} = -\beta \sum_i V(i)q_i$, comparing with 
Eq. (\ref{pf1}) now allows us to identify up to an overall constant $C$, the electrostatic potential as
\begin{equation}
V(i) = {i\over \beta q}(\phi(i) + C).
\end{equation}
Notably the potential drop across the system is given by
\begin{equation}
\Delta V = V(-1) - V(M)  
\end{equation}
where we include the sites $-1$ and $M$ which represent the capacitor plates. 

With electroneutrality, the integrand is invariant by a translation of the field $\phi(x)\rightarrow\phi(x)+\phi_0$. To keep a finite expression, we must set this constant enforcing, say, $\phi(0)=0$.
Inserting this expression in the grand partition function, we can easily perform the sum over the $S_j$ which are no longer coupled:
\begin{equation}
\Xi = \int_0^{2\pi} \frac{d\psi}{2\pi}\int [d\phi]\exp\left(-\frac{1}{2\gamma}\int\left[{d\phi(x)\over dx}\right]^2 dx\right)\prod_j(1+2\mu\cos(\phi(j)+\psi)) .
\end{equation}
For this path integral, where the field $\phi$ propagates freely between the integers, the propagator is known exactly
\begin{equation}
\int_{\phi(j)}^{\phi(j+1)} [d\phi]\exp\left(-\frac{1}{2\gamma}\int_{j}^{j+1}\left[{d\phi(x)\over dx}\right]^2 dx\right)=
{1\over{\sqrt{2\pi\gamma}}}\exp\left(-\frac{(\phi(j+1)-\phi(j))^2}{2\gamma}\right).
\end{equation}
This now gives
\begin{equation}
\Xi=\int_0^{2\pi} {d\psi\over 2\pi}\int \prod_j \frac{d\phi(j)}{\sqrt{2\pi\gamma}}\exp\left(-\sum_j\frac{(\phi(j+1)-\phi(j))^2}{2\gamma}+\sum_j\log(1+2\mu\cos(\phi(j) + \psi))\right).
\end{equation}
Finally we can rescale the variables $\phi(i) + \psi\to \phi(i)$ in the above integral to give
\begin{equation}\label{xi}
\Xi= {1\over 2\pi}\int \prod_j \frac{d\phi(j)}{\sqrt{2\pi\gamma}}\exp\left(-\sum_j\frac{(\phi(j+1)-\phi(j))^2}{2\gamma}+\sum_j\log(1+2\mu\cos(\phi(j)))\right).
\end{equation}
All integrals here are for $\phi(i)$ for $i\geq 1$ over the interval $(-\infty,\infty)$ but the integral 
over $\phi(0)$ is over $(-\pi,\pi)$.

\subsection{With charges on the boundaries or constant applied voltage}

Charges can be put directly on the boundaries of our system by inserting a charge $qQ$ on the site $-1$ and $-qQ$ on the site $M$. To obtain the partition function with these charges, we just have to add $iQ(\phi(-1)-\phi(M))$ in the exponential in (\ref{xi}) and extend the integration to over $\phi(-1)$
and $\phi(M)$, where we again restrain the integral over the field at the first site ($\phi(-1)$) which is now at $i=-1$ over $(-\pi,\pi)$ . This gives the grand partition function at fixed
external charges
\begin{equation}
\Xi_Q= {1\over 2\pi}\int \prod_j \frac{d\phi(j)}{\sqrt{2\pi\gamma}}\exp\left(-\sum_j\frac{(\phi(j+1)-\phi(j))^2}{2\gamma}+\sum_j\log(1+2\mu\cos(\phi(j))) + iQ(\phi(-1)-\phi(M)) \right).
\end{equation}
The grand potential is defined as 
\begin{equation}
\Omega_Q=-\beta^{-1}\log (\Xi_Q)=\beta^{-1}\omega_Q,
\end{equation}
with $\omega_Q$ being the dimensionless grand potential.

In this capacitor set up we will study the differential capacitance
\begin{equation}
C=q\frac{\partial Q}{\partial\Delta V},
\end{equation}
and for simplicity we will  work with dimensionless quantities: the reduced potential
\begin{equation}
v=\beta q V=-i\phi,
\end{equation}
and the reduced capacitance
\begin{equation}
c=\frac{\partial Q}{\partial\Delta v}.
\end{equation}

We can also impose a potential drop and look at its effect on the charges on the boundaries. To obtain the corresponding partition function, we have to restrict the integration in Eq. (\ref{xi}) only to be over fields satisfying 
\begin{equation}
\Delta V=\frac{i(\phi(M)-\phi(-1))}{\beta q},
\end{equation}
or in terms of the dimensionless quantities
\begin{equation}
\phi(-1)-\phi(M) = i\Delta v \label{dv}.
\end{equation}
We now note that this constraint can be imposed via a delta function written as a Fourier integral
\begin{equation}
\delta(\phi(-1)-\phi(M) - i\Delta v) = {1\over 2\pi} \int dQ \exp(iQ(\phi(-1)-\phi(M)-i\Delta v).
\end{equation}
The partition function and the grand potential at fixed potential difference can then be written, up to an unimportant factor of $2\pi$, as
\begin{equation}\label{xiv}
\Xi_{\Delta v} =\exp(-\omega_{\Delta v})= \int dQ\exp(Q\Delta v-\omega_Q).
\end{equation}
The interpretation of this equation is clear, the system is now free to accumulate charge at the plates 
in order to hold the potential constant. This result is in accordance with standard thermodynamic relations and is exploited in numerical simulations of systems with fixed voltage drops across them
\cite{kiyo2007}.

There is a clear difference between the two ensembles. In the constant charge ensemble the potential drop $\Delta v$ is a fluctuating quantity and from Eq. (\ref{dv}) its
average value is given as 
\begin{equation}
\langle \Delta v\rangle_Q  =\langle  -i (\phi(-1)-\phi(M))\rangle = {\partial\omega_Q\over \partial Q}.
\label{dvq}
\end{equation} 
However in the constant potential ensemble the surface charge $Q$ is a fluctuating quantity whose
average value is  given by
\begin{equation}
\langle Q\rangle_{\Delta v} = -{\partial\omega_{\Delta v}\over \partial \Delta v} .
\end{equation}
For each ensemble we have  a corresponding differential capacitance. At fixed $Q$ we have
\begin{equation}
c_{Q} = {\partial Q\over \partial \langle \Delta v\rangle_Q}
\end{equation}
and at fixed $\Delta v$ we have
\begin{equation}
c_{\Delta v} = {\partial \langle Q\rangle_{\Delta v}\over \partial \Delta v}.
\end{equation}
In this latter constant potential ensemble we have a simple fluctuation dissipation relation relating
the capacitance at constant voltage to the fluctuations of the charge on the capacitor plates:
\begin{equation}
c_{\Delta v} =  \langle Q^2\rangle_{\Delta v} -\langle Q\rangle^2_{\Delta v} \geq 0.
\end{equation}

We should note here that one of the first explanations of double layer capacitance is due to Helmholtz \cite{helm} who assumed that the  surface charge could be neutralized exactly by a layer of ions  at a distance $r$ from the surface, where $r$ is the effective ionic radius of the ions (assumed to be the same for both cations and anions). Applying this to our model (note the complete neutralization will not always be possible) we find a dimensionless Helmholtz capacitance 
\begin{equation}
c_{H} = {1\over 4\gamma}, \label{hc}
\end{equation}
where in this non-fluctuating case $c_Q$ and $c_{\Delta v}$ are the same.  This Helmholtz 
capacitance sets an intrinsic scale of capacitance in the model studied here.

\section{Mean field approximation}

In the mean field approximation, we compute the partition function by the saddle-point approximation, minimizing the argument of the exponential function. In Eq. (\ref{xi}), this gives the following equation for the potential (we write $v_j=v(j)$):
\begin{equation}\label{mfeq}
\frac{2v_j-v_{j+1}-v_{j-1}}{\gamma}+\frac{2\mu\sinh(v_j)}{1+2\mu\cosh(v_j)}=0.
\end{equation}
The reduced grand potential (measured in units of $k_B T$) is then given as
\begin{equation}
\omega=-\ln[\Xi]=\sum_j\frac{(v_{j+1}-v_j)^2}{2\gamma}-\sum_j\log(1+2\mu\cosh(v_j)).
\end{equation}
When the boundaries carry no charge or the applied potential $\Delta v = 0$, the solution to this equation is obvious: $v_j=0$. Then the pressure is easily obtained
\begin{equation}
p=\beta P=-\frac{\delta \omega_M }{\delta M}=\log(1+2\mu),
\end{equation}
where the above is a discrete derivative and the formula is valid in both the constant $Q$ and constant
$\Delta v$ ensembles. In this expression, we note that the charge carried by the particles does not enter in the pressure which is, of course, due to the mean-field approximation in the bulk.

In the case of a constant applied potential the boundary condition is simply given by $v_{-1}-v_M=\Delta v$. In the presence of charges on the boundary the fact that the system is globally electroneutral means that  the electric field is zero outside the system giving $v_0-v_{-1}=v_M-v_{M-1}=\gamma Q$.

Within the mean field approximation the local the charge density is given by 
\begin{equation}
\langle S_j\rangle=-\frac{2\mu\sinh(v_j)}{1+2\mu\cosh(v_j)}.
\end{equation}

There is a slight difference between this mean field approximation and that  found in \cite{Borukhov,Kornyshev1,Kornyshev2} where the continuous limit $a\rightarrow 0$ with the maximal density $q/a=\rho_0$ kept constant is taken. In this continuum approximation, in dimension-full
quantities the mean field equation becomes
\begin{equation}
\epsilon{d^2 V(x)\over dx^2}=\rho_0\frac{2\mu\sinh(\beta qV(x))}{1+2\mu\cosh(\beta qV(x))}.
\end{equation}
This limit is legitimate if the potential varies slowly compared to the length $a$, but is not correct in the general case \cite{Borukhov}. In order to compare the mean field theory with the full underlying model we  will numerically solve the finite difference equations to make as thorough a comparison possible between the mean field theory and our exact solution.

\section{Exact computation}

In what follows in order to lighten the notation we will denote $y_i=\phi(i)$. We define 
the operator 
\begin{equation}
p(y,y')=\frac{1}{\sqrt{2\pi\gamma}}\exp\left(-\frac{(y-y')^2}{2\gamma}\right)
\end{equation}
and also its operator square root
 \begin{equation}
 p^{1/2}(y,y')=\frac{1}{\sqrt{\pi\gamma}}\exp\left(-\frac{(y-y')^2}{\gamma}\right),
 \end{equation} 
 which obeys
 \begin{equation}
 \int dy' p^{1/2}(y,y')p^{1/2}(y',y'') = p(y,y'').
 \end{equation}
With these definitions  the grand partition function in the constant $Q$ ensemble can be written as
\begin{equation}\label{xi_ex}
\Xi_Q=\int dy_{-1} (1+2\mu\cos(y_0))\prod_{j=0}^{M} dy_j p(y_{j-1},y_j)(1+2\mu\cos(y_j)),
\exp(iQ(y_{-1}-y_{M}))
\end{equation}
where the integral over $y_{-1}$ is over $(-\pi,\pi)$ and the other integrals are over $(-\infty,\infty)$.
Mathematically it is now convenient to  introduce the symmetric operator
\begin{equation}\label{xi1}
K(y,y')=\int p^{1/2}(y,z)(1+2\mu\cos(z))p^{1/2}(z,y')dz,
\end{equation}
using which we may write
\begin{equation}
\Xi_Q ={1\over 2\pi} \int_{-\pi}^{\pi} dx \exp(iQx)\left[p^{1/2}K^Mp^{1/2}\right]\exp(-iQx),
\end{equation}
where we have used operator notation of the form
\begin{equation}
{\cal O}f(y) \equiv\int_{-\infty}^{\infty} {\cal O}(y,y')f(y') dy',
\end{equation}
for the operators ${\cal O}$ which are $K$ and $p^{1/2}$ in the above. Now, if we define the function
\begin{equation}
\Psi_Q(x) ={1\over \sqrt{2\pi}} p^{1/2}\exp(-iQx),
\end{equation}
we can write
\begin{equation}
\Xi_Q = \int_{-\pi}^{\pi} dx \ \Psi^*_Q (x) \left[K^{M} \Psi_Q\right](x) \equiv \langle \Psi_Q|K^M|\Psi_Q\rangle.
\end{equation}

The computation of the local dimensionless charge at the site $i$ is carried out as follows. In the carrying 
out the trace of the spins once decoupled we have 
\begin{equation}
\sum_{S_i =\pm 1} S_i \exp(i\phi_i S_i) = 2i\sin(\phi_i),
\end{equation} 
and we thus find that
\begin{equation}
\langle S_i\rangle = {\langle \Psi_Q|K^i L K^{M-i-1}|\Psi_Q\rangle\over \langle \Psi_Q|K^M|\Psi_Q\rangle},
\end{equation}
where 
\begin{equation}
L(y,y') = 2i \mu \int dz \ p^{1/2}(y,z)\sin(z)p^{1/2}(z,y').
\end{equation}

For practical computations it is best to analyze the action of the various operators $K$ and $L$ on
functions decomposed in term of Fourier modes. We define a $Q$-dependent class of functions since the vector space of functions having the Fourier representation
\begin{equation}
f(x) = \sum_{k\in \mathbb{Z}-Q} {\tilde f}_k \exp(ikx).
\end{equation}
We see that the classes where $Q$ have the same non-integer part are the same. The function $\Psi_Q$ is of $Q$ class as
\begin{equation}
p^{1/2}\exp(-iQx) = \exp\left(-{\gamma Q^2\over 4}\right)\exp(-iQx),
\end{equation}
and thus
\begin{equation}
\tilde \Psi_{Q,k} = {1\over \sqrt{2\pi}}\exp\left(-{\gamma Q^2\over 4}\right)\delta_{k,Q}.
\end{equation}
In this representation  we then have 
\begin{equation}
Kf(x)=\sum_k\widetilde{Kf}_k\exp(ikx),
\end{equation}
with 
\begin{equation}
\widetilde{Kf}_k = \exp(-\gamma k^2/4)\left(\exp(-\gamma k^2/4)\tilde f_k+\mu\left[\exp(-{\gamma (k-1)^2\over4})\tilde f_{k-1}+\exp(-{\gamma (k+1)^2\over 4})\tilde f_{k+1}\right]\right).
\end{equation}
Similarly in the same representation we have for the charge density operator $L$
\begin{equation}
\widetilde{Lf}_k = \mu \exp(-{\gamma k^2\over 4})\left(\exp(-{\gamma (k-1)^2\over 4}) \tilde f_{k-1}-\exp(-{\gamma (k+1)^2\over 4})\tilde f_{k+1}\right).
\end{equation}
In terms of matrix elements we can then write
\begin{eqnarray}
\tilde K_{k,l}&=&\exp(-{\gamma[k^2+l^2]\over 4})\left(\delta_{k,l}+\mu[\delta_{k,l-1}+\delta_{k,l+1}]\right),\label{modes}\\
\tilde L_{k,l}&=&\mu\exp(-{\gamma[k^2+l^2]\over 4})(\delta_{k,l+1} - \delta_{k,l-1}) .
\end{eqnarray}
The grand partition function now reads
\begin{equation}
\Xi_Q = \exp\left(-{\gamma Q^2\over 2}\right)\langle c_Q| \tilde K^M |c_Q\rangle,
\end{equation} 
where the surface charge vector $|c_Q\rangle$ has components $\tilde c_{Q,k} =\delta_{Q,k}$. The local charge density at site $i$ is then given by
\begin{equation}
\langle S_i\rangle = {\langle c_Q| \tilde K^i\tilde L \tilde K^{M-i-1}| c_Q\rangle\over \langle c_Q|  \tilde K^M | c_Q\rangle }.
\end{equation}
The above expressions are formally exact and are given in terms of infinite dimensional matrices. In practice we now need to numerically evaluate the expressions in order to obtain the exact thermodynamics of the model. In practice we see that modes with large $n$ are exponentially 
suppressed and thus we can truncate the matrices involved at sufficiently large $n$.

\subsection{Other formulation and limit of highly charged boundaries}

From another point of view the representation above presents a way of examining the problem in
the strong coupling limit as a discrete path integral. The matrix products involved can be expressed over
a discrete path integral of a discrete random walker $X_n$ which at each step changes by $\Delta X_n=\pm 1$ or $\Delta X_n=0$. This corresponds respectively to the presence of cations, anions or holes.
% (no particle respectively).
The grand potential at fixed charge $Q$ can be written as
\begin{equation}
\Xi_Q = \sum_{X_{-1}=0,X_{M}=0}\exp\left(-\sum_n{\gamma\over 2}(X_n-Q)^2 +\ln(\mu)\sum_n(X_n-X_{n+1})^2 \right),
\end{equation}
where the initial and final points of the process $X_n$ are fixed at zero. This representation is 
essentially a 1D Coulomb lattice gas version of the Villain representation for 
the 2D Coulomb gases \cite{villain, bask} and is also related to the electric field representation of the Coulomb gas used by Aizenman and Fr\"ohlich \cite{aiz}.  In the limit of large $\gamma$
it is clear that the process $X_n$ will be dominated by paths which start at zero and then stay as close 
as possible to the value $Q$. If we define by $\theta$ the maximal distance that $Q$ is from an integer 
we can write  $Q=n_Q+ \theta$ where by definition $\theta \in [-1/2,1/2]$ and $n_Q$ is integer.
For large $\gamma$ the process $X_n$ will increase from $X_{-1}=0$ to $X_{n_Q-1}=n_Q$ and 
similarly descend from this value when it is at distance $n_Q$ from the end of the path $m=N$. These initial and final parts of these paths have a contribution to the action $S$ of $\Xi_Q$ (where
$\Xi_Q = \exp(S)$) 
\begin{equation}
S_{\rm init}= -\gamma\sum_{n=0}^{n_Q} (Q-n)^2 + 2\ln(\mu) n_Q=-\gamma\sum_{n=0}^{n_Q}(n+|\theta|)^2 + 2\ln(\mu) n_Q.
\end{equation}
This first term is independent of $M$ and is not extensive in the system size, it can thus be identified
with a surface term in the free energy. Now, in the bulk there are two possibilities: the first is where
$\gamma \gg \ln(\mu)-1$ and in this case it is not energetically favorable to depart from the optimal 
value $n_Q$, the effective bulk action  thus being
\begin{equation}
S_{\rm bulk} = -{\gamma\over 2}(M-2n_Q)\theta^2.
\end{equation}
We thus see that in this limit the bulk free energy depends strongly on the surface charge. In the limit of large 
$\gamma$ but where $\ln(\mu) \gg {\gamma\over 2} \theta^2$, the system reduces its action by permitting charge oscillations about the optimal value $n_Q$, giving a bulk action of
\begin{equation}
S_{\rm bulk} \approx -{\gamma\over 4}(M-2n_Q) \left( \theta^2 + (1-|\theta|)^2\right]  + (M-2n_Q) \ln(\mu).
\end{equation}

This unusual dependence of the bulk thermodynamics on surface terms can be seen  
in an alternative manner. In the limit of large $M$ the thermodynamics in the constant $Q$ ensemble is dominated by the eigenvector $|\lambda_0(Q)\rangle$ of maximal eigenvalue $\lambda_0(Q)$ of the operator $\tilde K$ (thus $\tilde K|\lambda_0(Q)\rangle = \lambda_0(Q) |\lambda_0(Q)\rangle$). The partition function in the constant $Q$ ensemble obeys in this limit
\begin{equation}
\ln(\Xi_Q) = -{\gamma Q^2\over 2} + M\ln(\lambda_0(Q))  +\ln(\langle \lambda_0(Q)|c_Q\rangle^2).\label{tdl}
\end{equation}
The subtle point in Eq. (\ref{tdl}) is that the largest eigenvalue and eigenvector actually depend on 
$Q$. This phenomenon occurs in the ordinary continuum one dimensional Coulomb fluid without
hard core interactions, the thermodynamic limit thus depends on the boundary term. Physically this is related to the long range nature of the one dimensional Coulomb interaction and in terms of field theory
the ground states or $\theta$-vacua depend on the boundary terms \cite{aiz}. The system is thus sensitive to the  ability of the counter-ions to completely screen the surface charges and from our discussion above the eigenvectors and eigenvalues of $\tilde K$ depend on the non-integer part of $Q$. If one is in the strong coupling regime one may truncate the transfer matrix $\tilde K$ by keeping just the site $n_Q$ and  its neighbors. This gives the truncated matrix
\begin{equation}
\tilde K_{t} = \exp\left(-{\gamma\over 2}\theta^2\right)\begin{pmatrix}\exp\left(-{\gamma\over 2}(1+2|\theta|)\right) && \mu \exp\left(-{\gamma\over 4}(1+2|\theta|)\right) && 0 \cr \mu \exp\left(-{\gamma\over 4}(1+2|\theta|)\right) &&
1 && \mu \exp\left(-{\gamma\over 4}(1-2|\theta|)\right) \cr 0 && \mu \exp\left(-{\gamma\over 4}(1-2|\theta|)\right) &&  \exp\left(-{\gamma\over 2}(1-2|\theta|)\right)\end{pmatrix}.
\end{equation}
In the limit of finite $\mu$ but large $\gamma$, and when $\theta \neq 0$  (more precisely when $\ln(\mu)-1 \ll \gamma\theta$), we see that the matrix $\tilde K$ can be approximated as
\begin{equation}
\tilde K_{t} \approx \exp\left(-{\gamma\over 2}\theta^2\right)\begin{pmatrix}0&& 0&& 0 \cr 0 &&
1 && 0\cr 0 && 0 &&  0\end{pmatrix},
\end{equation}
and, in agreement with our argument above about dominant paths, we see that in this regime  we
have
\begin{equation}
\ln(\lambda_0(Q)) \approx -{\gamma\over 2}\theta^2.\label{plat1}
\end{equation}
This means that at relatively small values of $\mu$ integer values of the surface charge are
thermodynamically preferred and thus in the constant potential ensemble we should see plateaus
of roughly constant integer charge in the average surface charge as a function of applied voltage.  
However, still in the strong coupling limit of large $\gamma$ but close to full filling where $\mu\exp(-{\gamma/2}) \gg 1$, we have
\begin{equation}
\tilde K_{t} \approx \exp(-{\gamma\over 2}\theta^2)\begin{pmatrix}0&& \mu \exp\left(-{\gamma\over 4}(1+2|\theta|)\right) && 0 \cr \mu \exp\left(-{\gamma\over 4}(1+2|\theta|)\right) &&
0 && \mu \exp\left(-{\gamma\over 4}(1-2|\theta|)\right) \cr 0 && \mu \exp\left(-{\gamma\over 4}(1-2|\theta|)\right) && 0 \end{pmatrix},
\end{equation}
and for non-zero $\theta$ we thus obtain
\begin{equation}
\ln(\lambda_0(Q)) \approx -{\gamma \over 2}\left[  \left(|\theta| -{1\over 2}\right)^2 + {1\over 4} \right] + \ln(\mu),
\label{plat2}
\end{equation}
again in agreement with our discussion about dominant paths. Furthermore this shows that for large  $\mu$ and thus high filling or density, half integer values of the surface charge are
thermodynamically selected, and consequently the applied voltage vs. the average surface charge as a function of applied voltage should exhibit plateaus at half integer values of $Q$. 

Building on these arguments, we can evaluate the value of $\mu$ where the transition occurs. Let us compare the cases $\theta=0$ and $\theta=0.5$, choosing only the paths of maximal weight. When $\theta=0$, the path of maximal weight stays at $X_n=0$ and gives $\lambda_0(\theta=0)\simeq 1$. When $\theta=0.5$, the path of maximal weight oscillates freely between $X_n=1/2$ and $X_n=-1/2$, leading to the eigenvalue $\lambda_0(\theta=0.5)\simeq \exp(-\gamma/8)(\mu+1)$. The half-integer value is preferred when $\lambda_0(\theta=0.5)>\lambda_0(\theta=0)$, i.e. when
\begin{equation}
\label{approx_trans}
\mu>\exp(\gamma/8)-1.
\end{equation}
This approximate result will be compared with numerical computations later.

In the fixed $Q$ ensemble our approximate results and Eq. (\ref{dvq}) yield the following results
for the voltage drop in the strong coupling and thermodynamic limit
\begin{eqnarray}
\langle \Delta v\rangle_Q &=& M\gamma \theta \ \ \ {\rm for} \  \  \ \gamma \theta^2 \gg \ln(\mu) \\
&=& M\gamma(\theta -{1\over 2}{\rm sign}(\theta)) \ \ \ {\rm for}\ \ \  \gamma \theta^2 \ll \ln(\mu).
\end{eqnarray}
Note that when $\theta$ is small and $M$ is finite, the boundary terms coming from $S_{\rm init}$ will
dominate and the potential drop will only be weakly dependent on the system size. Interestingly we see that this extensive contribution to the free energy oscillates. In the limit where there are few ions 
(thus small $\mu$) we see that the potential drop has the same sign as $\theta$. However, when there 
are many ions (large $\mu$) the potential drop  has a sign opposite to that of $\theta$. 

We emphasize that these results depend on the assumption that $\gamma$ is small, for instance at high temperature. The oscillatory effects seen here should weaken as the temperature is increased, however their underlying presence should remain visible.  We also note that they should vanish when $\theta$ is
zero (i.e. for integer $Q$) but should be maximal when $\theta=\pm 1/2$. 

In the constant voltage ensemble the system can minimize its bulk electrostatic energy by tuning the surface charge to be and integer, and thus $\theta = 0$; this observation thus predicts the appearance
of a plateau like behavior in the charge as a function of the applied potential drop $\Delta v$.
The effective action in the constant $V$ ensemble is then
\begin{equation}
S = -\gamma\sum_{n=0,Q} n^2 + 2(\ln(\mu)-1)Q + Q\Delta v = -{\gamma\over 6}(2Q+1)(Q+1)Q + 2(\ln(\mu)-1)Q + Q\Delta v .
\end{equation}
In the limit of large $\Delta v$, minimization of the above action then leads to  the scaling law $\langle Q\rangle_{\Delta v} \sim |\Delta v|^{1\over 2}/\gamma$ and thus the capacitance behaves as
\begin{equation}
c_{\Delta v} \sim {1\over 2\gamma {|\Delta v|}^{1\over 2}},
\end{equation}
in accordance with mean field theory predictions \cite{Kornyshev1}. 

\section{Numerical computation}

We have, in principle, a formal solution to the thermodynamics of the LCF capacitor and explicit analytic results can be obtained in certain regimes. In general, however,  the partition function needs to be computed numerically. This is quite straightforward as matrix elements  $\tilde K_{k,l}$  decay exponentially when $k$ and $l$ are large. It is nevertheless important to keep matrix elements around  $n$, $m$ zero, as these are necessary in order to interact with the boundary term. We must thus keep modes $k\in\left[-(N-1)/2-Q,(N+1)/2-Q\right]$, where the $N$ is an odd  integer. The criteria that the matrix elements neglected by  this truncation are small, means that we must choose $\exp(-\gamma k^2/ 4)$ to be small when $k= \pm (N-1)/2-Q$. Practically the numerical evaluation is extremely fast and the dependence on the truncation value $N$ can be eliminated by increasing $N$ until no difference in the numerical results is seen.

In the following numerical results, almost all the parameters are set to 1, except if the effect of their variation is studied. We thus have $a=1$, $A=1$ $q=1$, $\beta=1$, $\epsilon=1$, $\mu=1$. With this choice of parameters, taking $\mu=1$ corresponds roughly to half-filling).

\subsection{Free energy for charged boundaries}

\begin{figure}
 \begin{center}
\includegraphics[angle=0,width=0.5\linewidth]{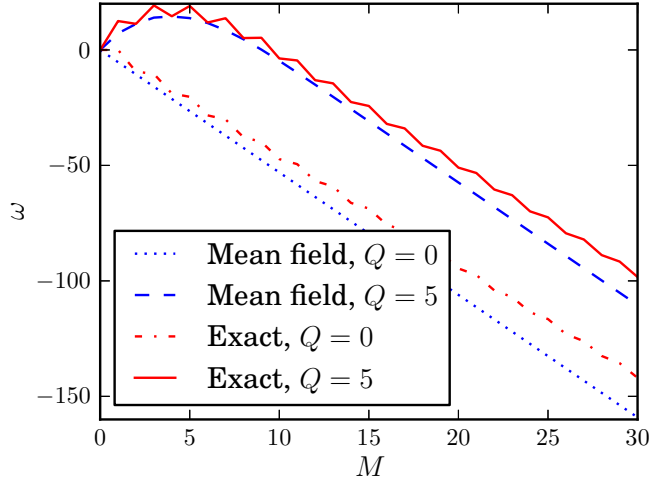}
\end{center}
\caption{Dimensionless grand potential as a function of the system size for $\gamma=1$ and $\mu=100$.}
\label{f_M_om}
\end{figure}

\begin{figure}
 \begin{center}
\includegraphics[angle=0,width=0.5\linewidth]{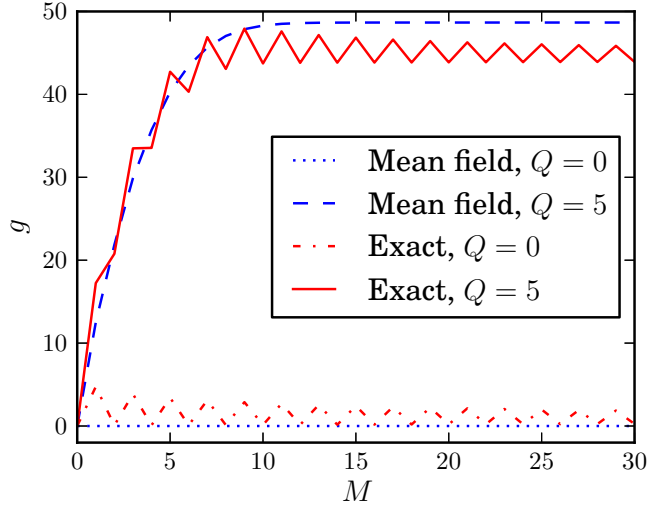}
\end{center}
\caption{Dimensionless free enthalpy as a function of the system size for $\gamma=1$ and $\mu=100$.}
\label{f_M_g}
\end{figure}

In Fig. \ref{f_M_om} we plot the reduced grand potential $\omega_Q$ as a function of the size of the system for two values of the charge fixed at the boundaries, for the exact solution and the mean field approximation. The dimensionless pressure is defined as $p_{M+\frac{1}{2}}=\omega_M-\omega_{M+1}$. We observe that
\begin{itemize}
\item The free energy quickly turns linear. This asymptotic regime is reached for $M\sim 2Q$, 
implying that the structures of the surface layers do not significantly  change once the system contains enough charge to screen the two charged boundary plates. In this regime of large plate separations, the pressure converges to a constant, which is the bulk pressure, $p\ind{b}=\lim_{M\rightarrow\infty}p_{M+1/2}$. We can then define the free enthalpy,
\begin{equation}
g=\omega+Mp\ind{b},
\end{equation}
which is presented in Fig. \ref{f_M_g}; its derivative with respect to plate separation gives the disjoining pressure, $p\ind{d}=p-p\ind{b}$.
\item The bulk pressure depends on the boundary charge with periodicity one (in accordance with our analysis above). However the dependence is rather weak and can only be seen in systems containing more than $10^4$ sites.
\item The interaction between the boundary plates is attractive when the separation is small, becoming repulsive when the size becomes larger than $\sim 2Q$. This is physically easy to explain: at large separations, the charges are completely screened by the ions. However, as in dilute electrolyte systems,
there is a slight excess osmotic pressure due to the screening layers that are drawn into the system which gives a (weak) positive pressure at large plate separations. On the other hand, when the plates are close to each other, there is not enough space for ions to enter and screen them, they thus attract each other simply because of their opposite charge. The disjoining pressure goes to zero at large distances.
\item We find that  that having an even number of sites is thermodynamically favorable with respect to an odd number of sites, this behavior is smoothed out as the size increases and is not seen in the mean field approximation. The physical interpretation is very simple: systems with an odd number of sites cannot be completely  filled, because of the electroneutrality condition, that induces a high free energy cost when $\mu$ is large ($\mu=100$ in these plots). When the system becomes large, this effect is not important, because full configurations do not dominate the partition function.
\item On the mean field level, the general shape of the grand potential remains unchanged, but the oscillations between even and odd numbers of sites disappears. The bulk pressure is also different: it is slightly higher in the mean field approximation.
\end{itemize}

We note here that the bulk pressure is a fundamental quantity that can be computed directly from the largest eigenvalue of the transfer matrix $K$,
\begin{equation}
p\ind{b}=\log(\lambda_0).
\end{equation}

\subsection{Average potential drop for fixed surface charges}

\begin{figure}
 \begin{center}
\includegraphics[angle=0,width=0.5\linewidth]{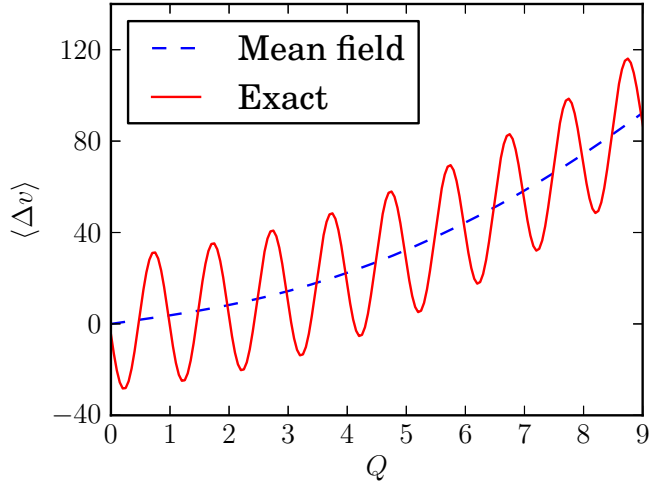}
\end{center}
\caption{Average potential drop as a function of the imposed charge for $\gamma=1$ and $\mu=1$ and system size $10^4$.}
\label{f_Q_V}
\end{figure}

Here we compute numerically $\langle \Delta v\rangle$ for fixed charge $Q$ for a large system ($10^4$ for the exact computation and $10^3$ for the mean field approximation). Our results are shown in Fig. \ref{f_Q_V}.

The exact result can be split into two parts: a monotonic part almost equal to the mean field result, and an oscillating part. As predicted by our analytical treatment the oscillating part is extensive. The mean field result and the monotonic part of the exact result do not depend on the size of the system if it is large enough.

The mean field result is again simple to explain: the surface charges are screened by a layer of cations and anions localized near the electrodes, in which the voltage drop takes place. These layers screen the surface charges so that the  charge due to each plate seen by the bulk is zero, and thus the bulk electric field vanishes. This explains the absence of an extensive part in the potential drop. The oscillating, extensive part in the exact result means that the charges are not perfectly screened. As $Q$ varies, the effective charge $Q\ind{eff}(Q)$, corresponding to the imperfect screening, oscillates with period 1, as predicted by our analytical results above.

\subsection{Average charge for fixed potential difference}

\begin{figure*}[t!]\begin{center}
	\begin{minipage}[b]{0.485\textwidth}
	\begin{center}
		\includegraphics[width=\textwidth]{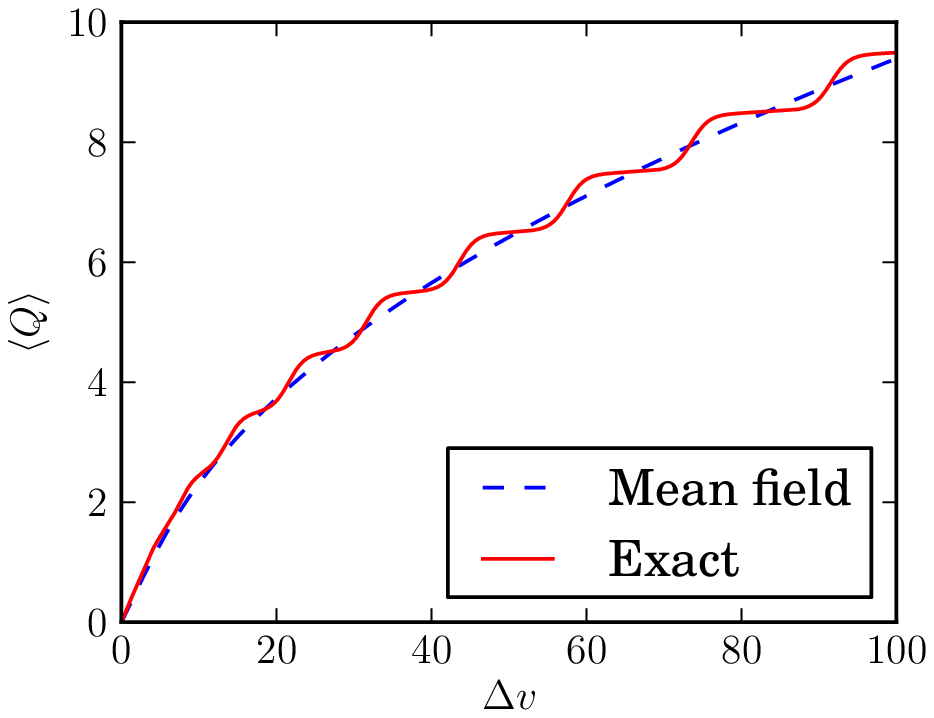}
	\end{center}\end{minipage} \hskip0.25cm
	\begin{minipage}[b]{0.485\textwidth}\begin{center}
		\includegraphics[width=\textwidth]{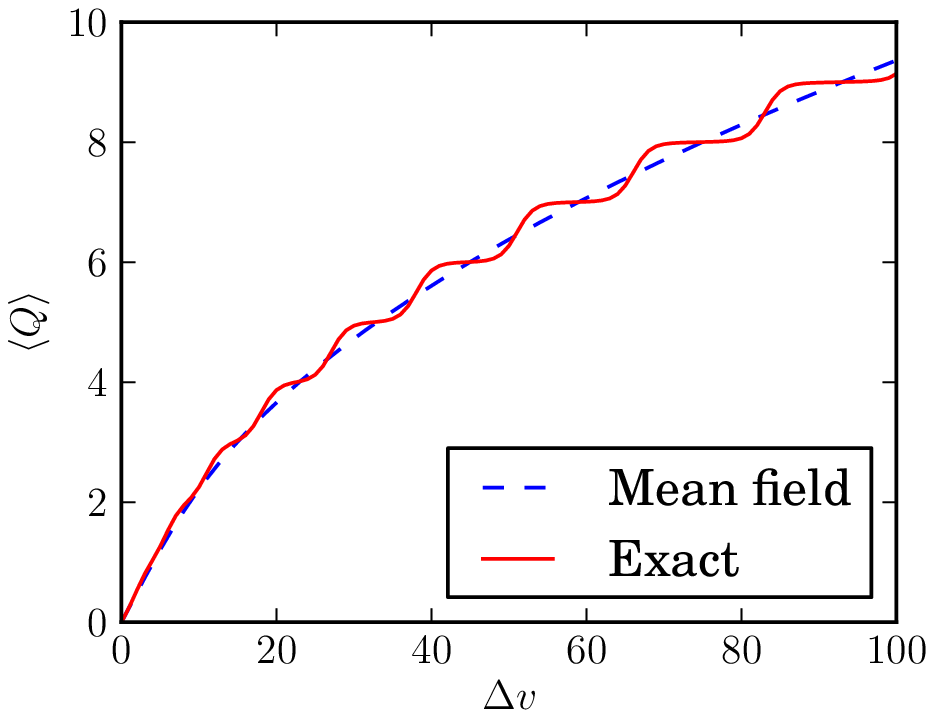}
	\end{center}\end{minipage} \hskip0.25cm	
\caption{Left: Average surface charge as a function of the imposed voltage, for $\gamma= 1$ and $\mu=1$ with system size $10^4$. Right: Average surface charge as a function of the imposed voltage, for $\gamma= 1$ and $\mu=0.5$ with system size $10^4$.}
\label{f_V_Q}
\end{center}\end{figure*}

The average boundary charge is shown as a function of the applied  voltage in Fig. \ref{f_V_Q}. The exact result increases in the same way as the mean field result, but when the voltage becomes large, plateaus are observed in the average charge, located at half-integer or integer values, (Fig. \ref{f_V_Q}),  depending on the parameters $\gamma$ and $\mu$. The plateaus show up only for large systems and they are smoothed out for small systems. 

%\begin{figure}
% \begin{center}
%\includegraphics[angle=0,width=0.5\linewidth]{V_Q.eps}
%\end{center}
%\caption{Average surface charge as a function of the imposed voltage, for $\gamma= 1$ and $\mu=1$ with system size $10^4$. }
%\label{f_V_Q}
%\end{figure} 

\begin{figure}
 \begin{center}
\includegraphics[angle=0,width=0.7\linewidth]{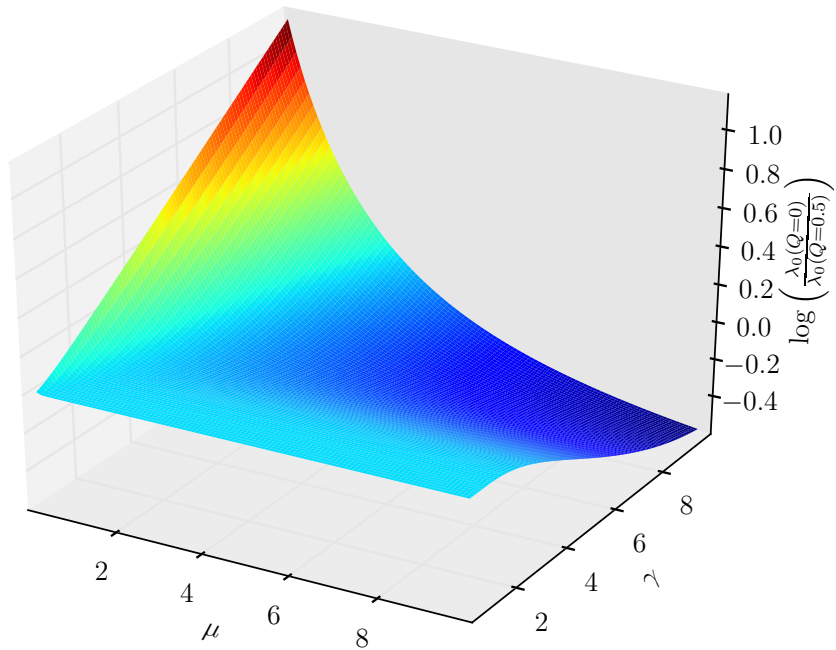}
\end{center}
\caption{Difference of the bulk pressure as a function of the parameters $\gamma$ and $\mu$. Contour line where $\lambda_0(Q=0)=\lambda_0(Q=0.5)$ (blue solid line) compared with the approximate transition line (\ref{approx_trans}) (red dashed line).}
\label{f_dvp}
\end{figure}

The change in the plateau locations is due to the fact that integer surface charges are more stable at low $\mu$, whereas half-integer charges are favored at high $\mu$ as we have shown in the large $\gamma$ limit in Eqs. (\ref{plat1}) and (\ref{plat2}), which show explicitly that the largest eigenvalue of the transfer matrix, and thus the bulk pressure, depends on $Q$. We have shown the difference of the bulk pressure between $Q=0$ and $Q=0.5$ in Fig. (\ref{f_dvp}), and compared it to the approximate transition line (\ref{approx_trans}). The surface charge $Q=0$ is favored for small values of $\mu$ (dilute regime) and the surface charge $Q=0.5$ is favored for large values of $\mu$ (dense regime).  This change in the plateau positions has an interesting effect on the capacitance near the point of zero charge and is addressed in the next section. 

%\begin{figure}
% \begin{center}
%\includegraphics[angle=0,width=0.5\linewidth]{V_Q2.eps}
%\end{center}
%\caption{Average surface charge as a function of the imposed voltage, for $\gamma= 1$ and $\mu=0.5$ with system size $10^4$.}
%\label{f_V_Q2}
%\end{figure} 

\subsection{Capacitance}

We computed the capacitance as a function of the voltage for a large system ($M=10^4$) and for different values of the parameters $\mu$ and $\gamma$. Two phenomena appear in the behavior of the capacitance:
\begin{itemize}
\item the capacitance can show multiple oscillations which increase in number as the coupling $\gamma$ is increased and as the applied voltage $\Delta v$ is increased.
\item These oscillations seem to be superimposed upon a trend predicted by mean-field theory with 
underlying bell and camel-like shapes, the former with a maximum at zero applied voltage and the latter
exhibiting a minimum. 
\end{itemize}

\begin{figure*}[t!]\begin{center}
	\begin{minipage}[b]{0.485\textwidth}
	\begin{center}
		\includegraphics[width=\textwidth]{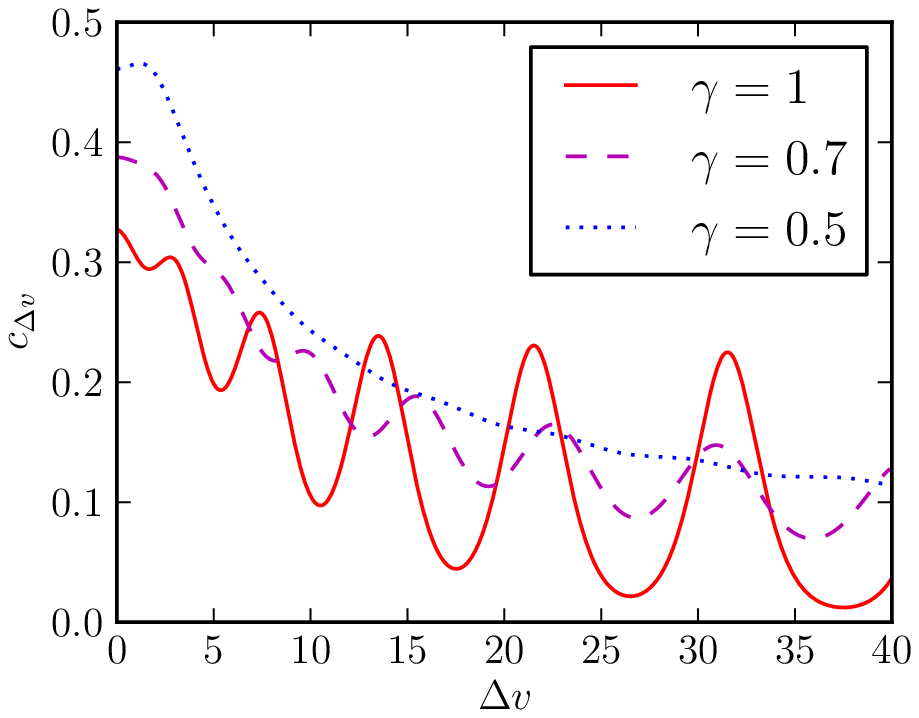}
	\end{center}\end{minipage} \hskip0.25cm
	\begin{minipage}[b]{0.485\textwidth}\begin{center}
		\includegraphics[width=\textwidth]{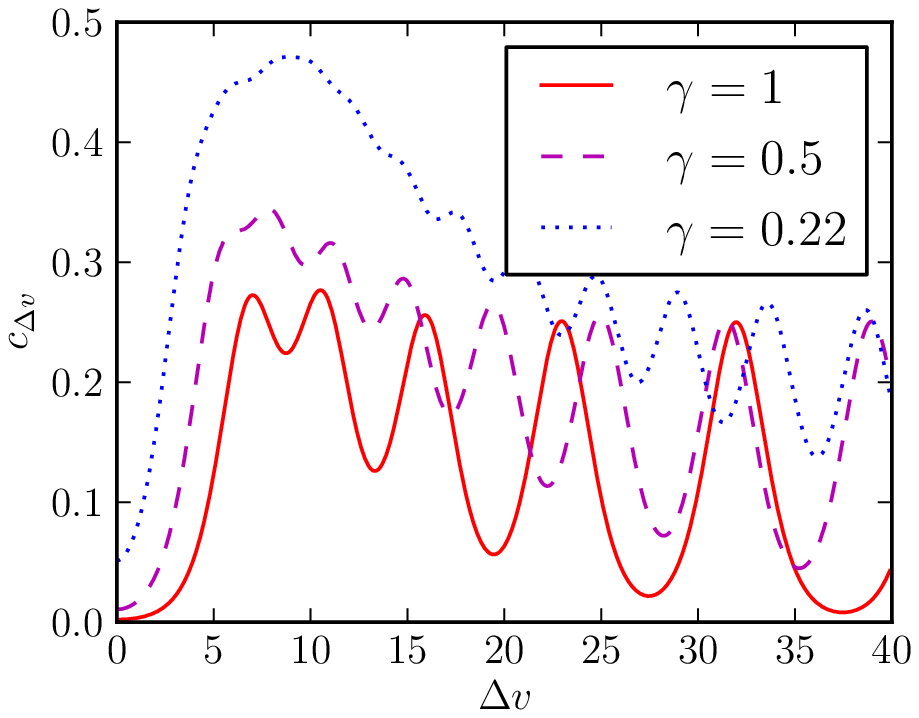}
	\end{center}\end{minipage} \hskip0.25cm	
\caption{Left: Capacitance as a function of the voltage drop for $\mu=1$: appearance of the peaks as $\gamma$ increases for a bell shaped capacitance. Right: Capacitance as a function of the voltage drop for $\mu=0.03$: appearance of the peaks as $\gamma$ increases for a camel shaped capacitance.}
\label{f_V_C_trans_bell}
\end{center}\end{figure*}

Examples are shown on Fig. \ref{f_V_C_trans_bell}. We see that the peaks appear at low temperatures (high $\gamma$), and the camel-like peak appears at low densities (low $\mu$) in accordance with mean-field theory \cite{Kornyshev1}.  The peaks are related to the plateaus in the average charge, so they are reduced when the system gets smaller.

\begin{figure*}[t!]\begin{center}
	\begin{minipage}[b]{0.485\textwidth}
	\begin{center}
		\includegraphics[width=\textwidth]{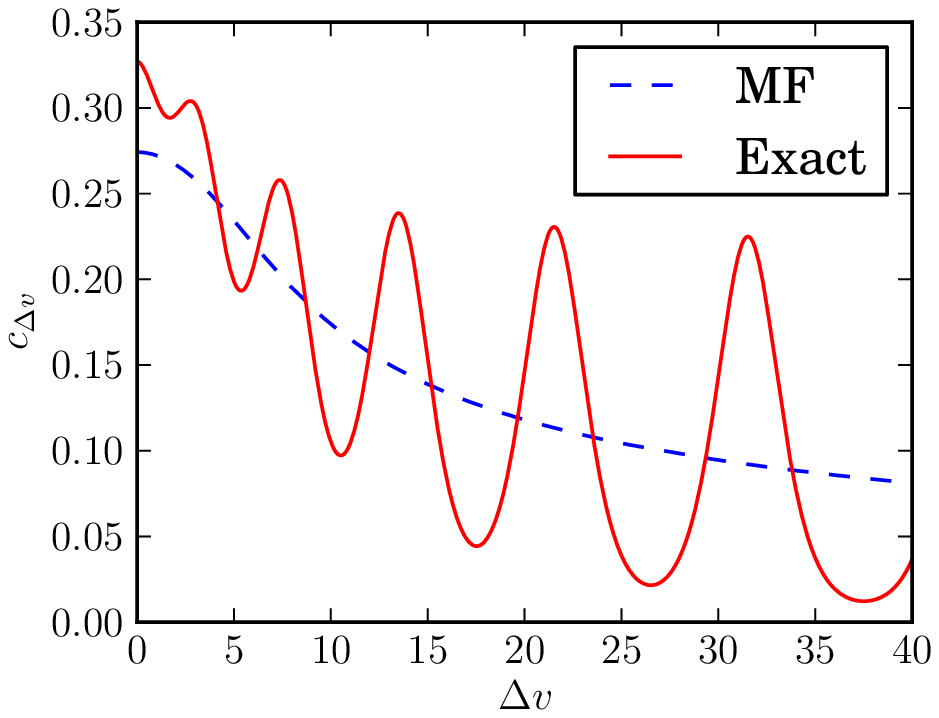}
	\end{center}\end{minipage} \hskip0.25cm
	\begin{minipage}[b]{0.485\textwidth}\begin{center}
		\includegraphics[width=\textwidth]{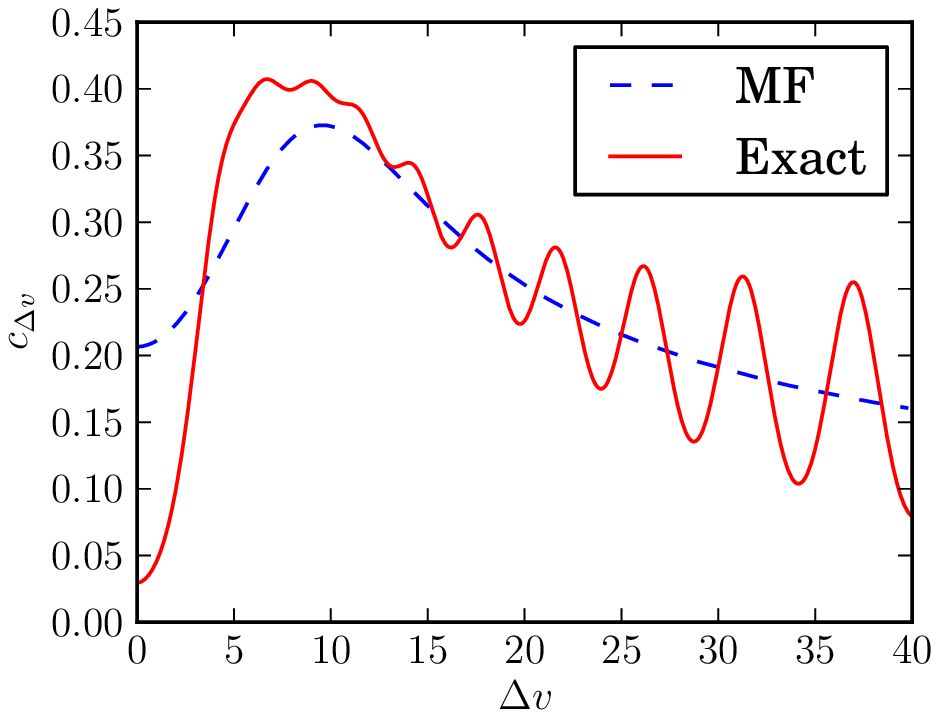}
	\end{center}\end{minipage} \hskip0.25cm	
\caption{Left: Capacitance as a function of the voltage drop for $\mu=1$ and $\gamma=1$. Right: Capacitance as a function of the voltage drop for $\mu=0.03$ and $\gamma=0.3$.}
\label{f_V_C_bell}
\end{center}\end{figure*}

\begin{figure}
 \begin{center}
\includegraphics[angle=0,width=0.5\linewidth]{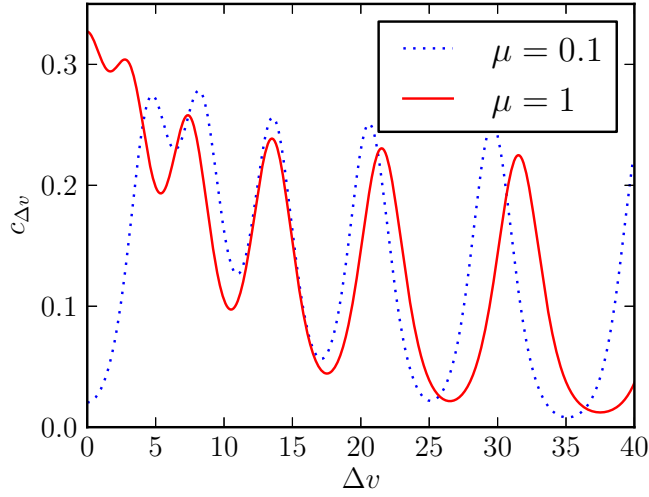}
\end{center}
\caption{Capacitance as a function of the voltage drop for $\gamma=1$ and and different values of the fugacity.}
\label{f_V_C_trans_mu}
\end{figure}

The exact results are compared to the mean field results for the bell and camel (Fig. \ref{f_V_C_bell}) shaped capacitance. The mean-field capacitance exhibits the same behavior, without the peaks, which is coherent with the absence of plateaus. They also strongly resemble the capacitance appearing in \cite{Kornyshev1}. We also plot the capacitance curve when $\mu$ increases, for $\gamma=1$, on Fig. \ref{f_V_C_trans_mu}, which shows the transition between the camel shape and the bell shape.

\begin{figure}
 \begin{center}
\includegraphics[angle=0,width=0.5\linewidth]{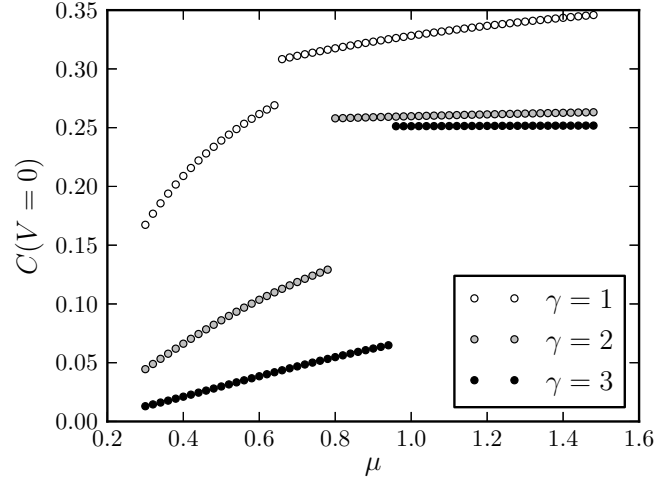}
\end{center}
\caption{PZC capacitance as a function of $\mu$ for different values of $\gamma$, for an infinite system.}
\label{f_mu_C}
\end{figure}

We now discuss the implication of the change in the plateau positions. When they are located at integer charges, the PZC, $Q=0$, corresponds to  a plateau; the surface charge will thus respond very slowly to changes in the imposed voltage and the capacitance is expected to be small. On the other hand, when the plateaus are located at half integer charges, the PZC lies between two plateaus and the capacitance is expected to be large. In the thermodynamic limit, when the system size goes to infinity, we thus may even expect a sharp phase transition. We plot the capacitance at the PZC as a function of $\mu$ in Fig. \ref{f_mu_C}; we indeed see a phase transition, which becomes sharper as temperature decreases.

A final point that should be mentioned is that the capacitance in the exact solution can {\sl exceed} the Helmholtz capacitance $c_H$.

%\begin{figure}
% \begin{center}
%\includegraphics[angle=0,width=0.5\linewidth]{V_C_trans_bell.eps}
%\end{center}
%\caption{Capacitance as a function of the voltage drop for $\mu=1$: appearance of the peaks as $\gamma$ increases for a bell shaped capacitance.}
%\label{f_V_C_trans_bell}
%\end{figure} 

%\begin{figure}
% \begin{center}
%\includegraphics[angle=0,width=0.5\linewidth]{V_C_trans_camel.eps}
%\end{center}
%\caption{Capacitance as a function of the voltage drop for $\mu=0.03$: appearance of the peaks as $\gamma$ increases for a camel shaped capacitance.}
%\label{f_V_C_trans_camel}
%\end{figure} 

%\begin{figure}
% \begin{center}
%\includegraphics[angle=0,width=0.5\linewidth]{V_C_bell.eps}
%\end{center}
%\caption{Capacitance as a function of the voltage drop for $\mu=1$ and $\gamma=1$.}
%\label{f_V_C_bell}
%\end{figure} 

%\begin{figure}
% \begin{center}
%\includegraphics[angle=0,width=0.5\linewidth]{V_C_camel.eps}
%\end{center}
%\caption{Capacitance as a function of the voltage drop for $\mu=0.03$ and $\gamma=0.3$.}
%\label{f_V_C_camel}
%\end{figure} 

\subsection{Charge density}

\begin{figure*}[t!]\begin{center}
	\begin{minipage}[b]{0.485\textwidth}
	\begin{center}
		\includegraphics[width=\textwidth]{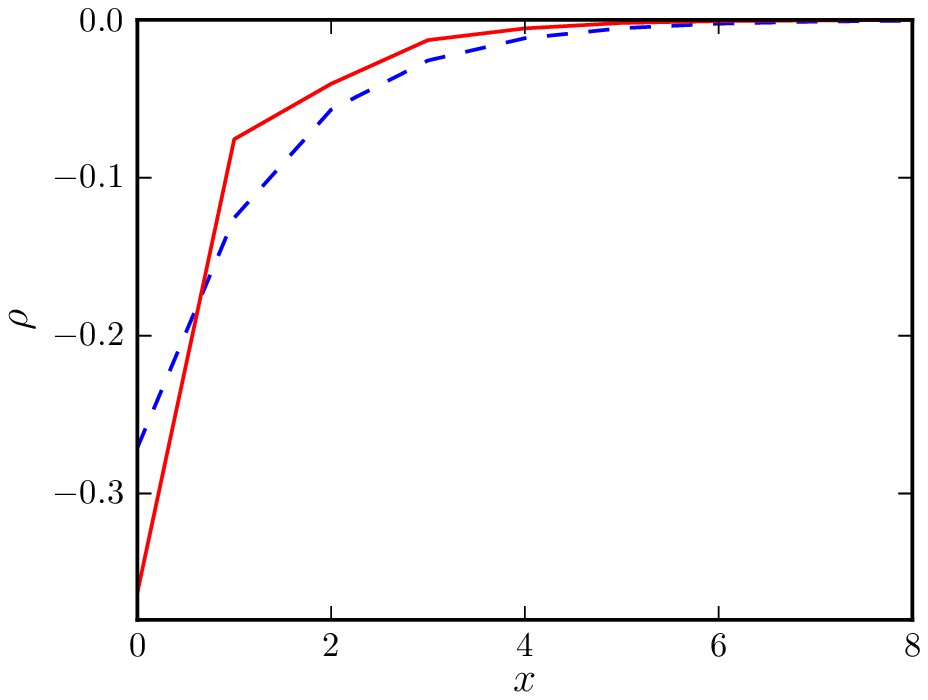}
	\end{center}\end{minipage} \hskip0.25cm
	\begin{minipage}[b]{0.485\textwidth}\begin{center}
		\includegraphics[width=\textwidth]{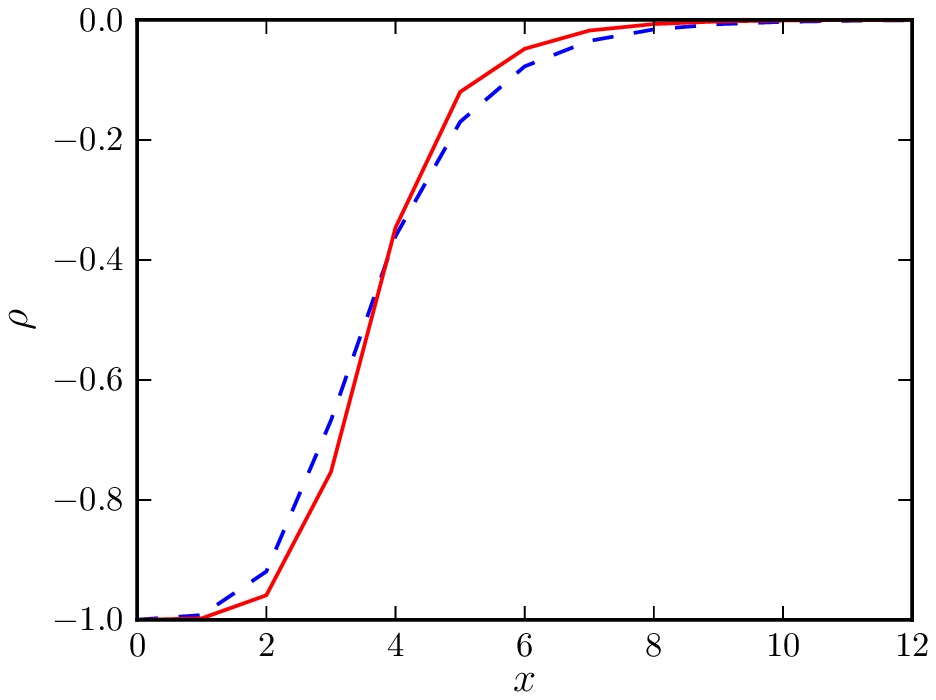}
	\end{center}\end{minipage} \hskip0.25cm	
\caption{Left: Charge density close to the left electrode (located at $x=-1$) as a function of the position for $\gamma=1$, $\mu=1$ and $Q=1$: exact result (solid line) and mean field result (dashed line). Right: Charge density close to the left electrode (located at $x=-1$) as a function of the position for $\gamma=1$, $\mu=1$ and $Q=4.25$: exact result (solid line) and mean field result (dashed line).}
\label{f_x_rho}
\end{center}\end{figure*}

The charge density versus the position is plotted on Fig. \ref{f_x_rho}; we show only the charge density close to the left boundary, the same profile with opposite charge is found close to the right boundary. It shows a counter-ion layer with a size corresponding to the charge $Q$. 

\begin{figure*}[t!]\begin{center}
	\begin{minipage}[b]{0.485\textwidth}
	\begin{center}
		\includegraphics[width=\textwidth]{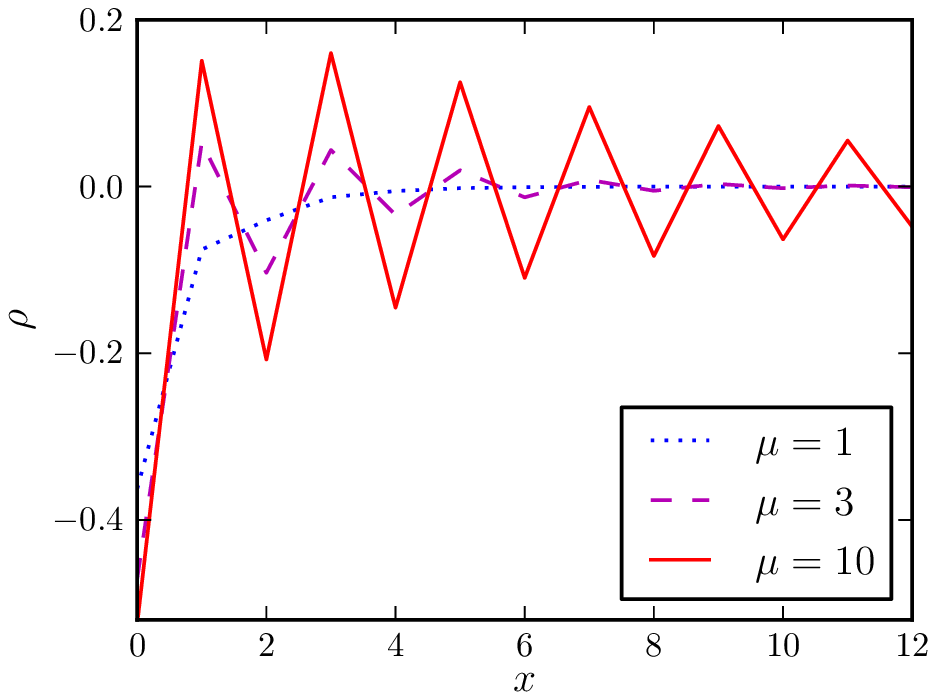}
	\end{center}\end{minipage} \hskip0.25cm
	\begin{minipage}[b]{0.485\textwidth}\begin{center}
		\includegraphics[width=\textwidth]{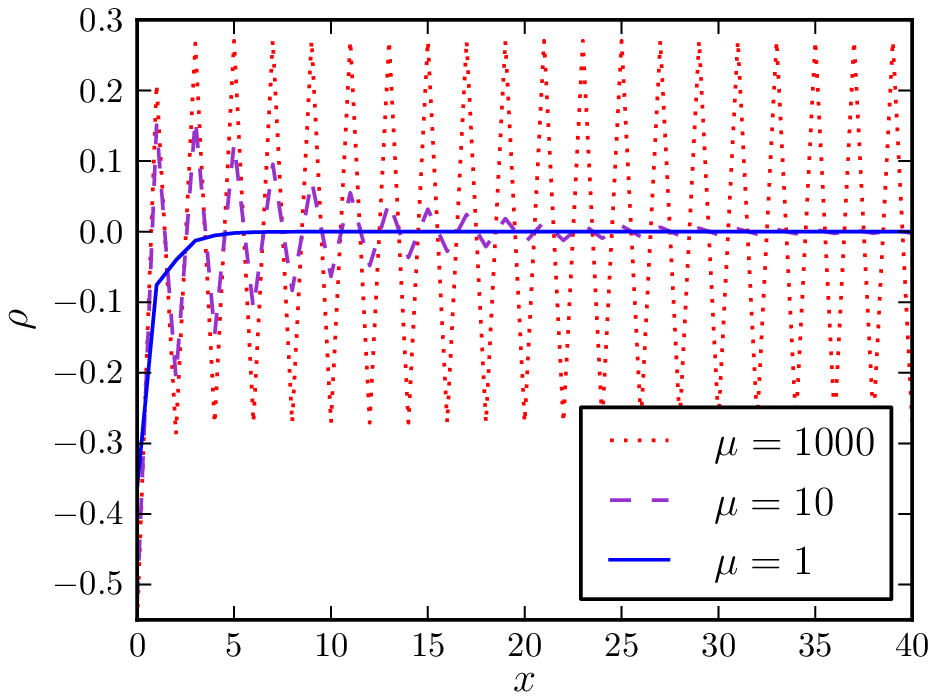}
	\end{center}\end{minipage} \hskip0.25cm	
\caption{Left: Exact result for the mean charge density close to the left electrode (located at $x=-1$) as a function of the position for $\gamma=1$, $\mu=1$ and $Q=0.5$, for different values of the fugacity. Right: Exact result for the mean charge density as a function of the position for $\gamma=1$, $\mu=1$ and $Q=0.5$, for different values of the fugacity.}
\label{f_x_rho_layers}
\end{center}\end{figure*}

\begin{figure}
 \begin{center}
\includegraphics[angle=0,width=0.5\linewidth]{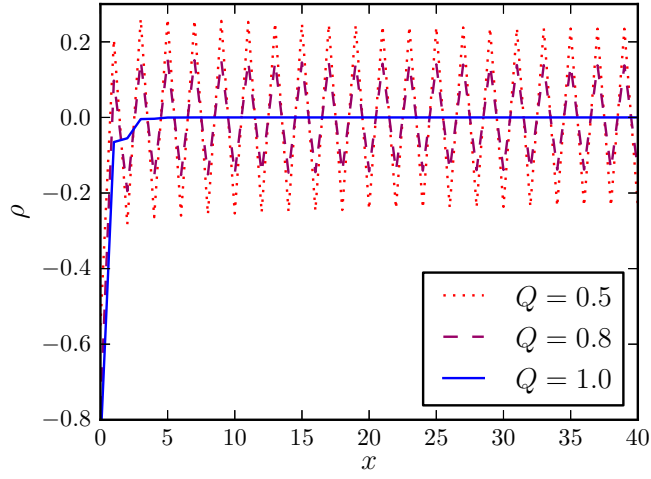}
\end{center}
\caption{Exact result for the mean charge density as a function of the position for $\gamma=1$ and $\mu=100$, for different values of the boundary charge.}
\label{f_x_rho_layers_q}
\end{figure}

\begin{figure}
 \begin{center}
\includegraphics[angle=0,width=0.5\linewidth]{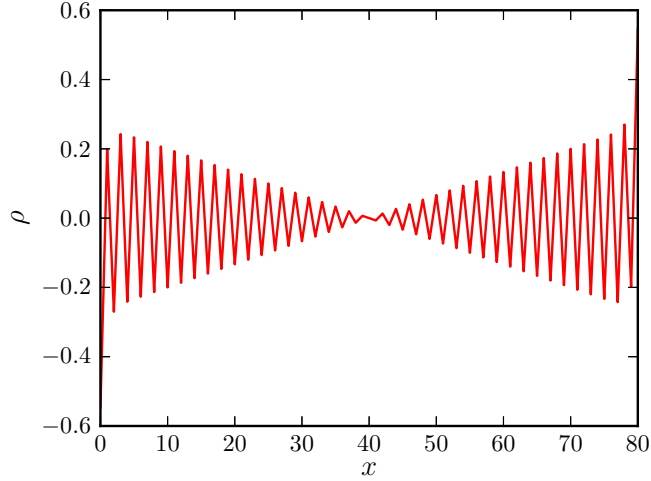}
\end{center}
\caption{Exact result for the mean charge density as a function of the position for $\mu=1000$ and $Q=0.5$, for an odd number of sites ($M=81$).}
\label{f_x_rho_layers_odd}
\end{figure}

If we increase the density ($\mu>1$) for a non integer charge $Q$, the charge density shows oscillations, as is seen on Fig. \ref{f_x_rho_layers}, corresponding to the ion layers as observed in \cite{Mezger}. The  phenomenology exhibited is:
\begin{itemize}
\item The amplitude of the oscillations varies continuously with the charge: it is maximal for half-integer charges and zero for integer charges, which is physical because an integer charge can be screened perfectly, cf. Fig. \ref{f_x_rho_layers_q}.
\item  The mean field result does not show oscillations (as can be seen in \cite{Borukhov,Kornyshev1}): they are thus a basic outcome of the exact result. 
\item We show these oscillations on a larger length scale in Fig. \ref{f_x_rho_layers}. The oscillations are damped with a characteristic length $\xi$, which increases with $\mu$, and we notice that $\xi\sim\mu$. Actually, the two highest eigenvalues of the transfer matrix $K$, $\lambda_0$ and $\lambda_1$, define a correlation length via $\xi=\left(\ln \left|\lambda_0/\lambda_1\right|\right)^{-1}$. An analytic computation for three Fourier modes and $Q=0$, and for two modes and $Q=0.5$, indeed gives $\xi\sim\mu$. 
\item The physical idea is that each boundary (left and right) creates such oscillations. On the plots described above, these oscillations sum up, because there is an even number of sites in the system. If there is an odd number of sites, the sum of the oscillations coming from the left and right boundary is destructive, as shown in Fig. \ref{f_x_rho_layers_odd}. This effect is obviously irrelevant for large (experimental) systems where $M\gg\xi$.
\end{itemize}

We also note that if the charge on the boundary is small, the ionic densities do not saturate, which is not in agreement with what is found in \cite{Borukhov} on the continuous mean field level.

It can also be seen in the exact result that  the surface charges are not necessarily perfectly screened (on average) by the ions. This can be established by integrating the average charge in the fluid over the first half of the space (which contains the left layer) which does not always give $-Q$. The average screening is perfect only when $Q$ is integer or half-integer.

\begin{figure}
 \begin{center}
\includegraphics[angle=0,width=0.5\linewidth]{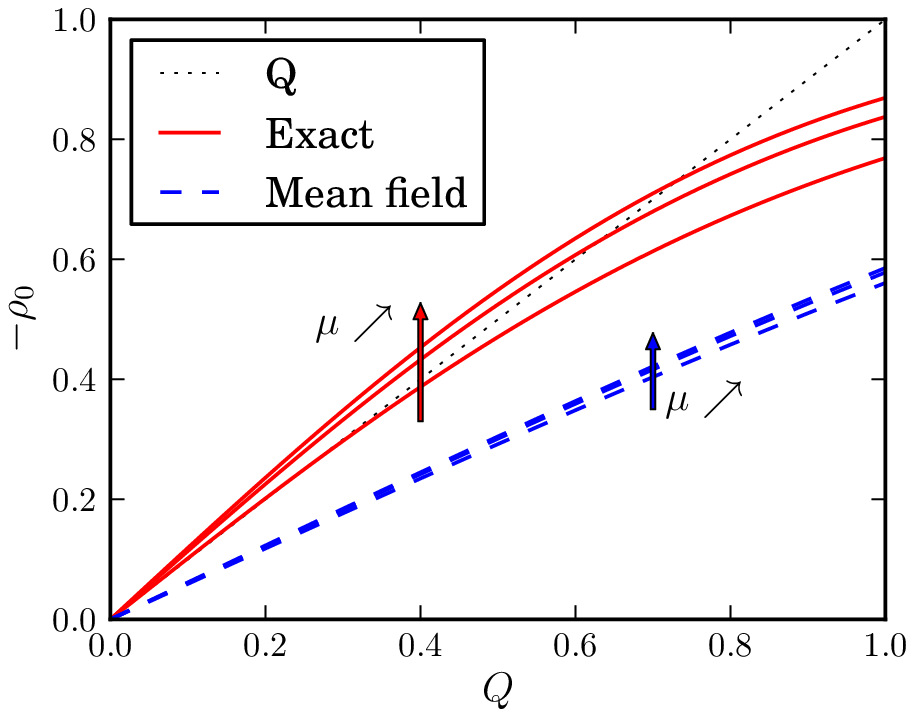}
\end{center}
\caption{Mean charge density $\rho_0$ of the first layer as a function of the surface charge $Q$ for $\gamma=1$ and $\mu\in\{3,10,100\}$: exact result (solid line) and mean field approximation (dashed line).}
\label{f_rho_overscreen}
\end{figure}

The phenomenon of over-screening cab be seen in  Fig. \ref{f_rho_overscreen}. In the exact treatment, and for high densities, the charge of the first layer may be higher than the boundary charge, and thus over-screens it. However, no over-screening can be seen in the mean field approximation. We should also emphasize that over-screening is {\sl not} a necessary condition to observe charge oscillations, e.g. oscillations exist for $\gamma=1$ and $\mu=3$ (cf. Fig. \ref{f_x_rho_layers}), a case where there is no over-screening.

%\begin{figure}
% \begin{center}
%\includegraphics[angle=0,width=0.5\linewidth]{x_rho2.eps}
%\end{center}
%\caption{Charge density close to the left electrode (located at $x=-1$) as a function of the position for $\gamma=1$, $\mu=1$ and $Q=1$: exact result (solid line) and mean field result (dashed line).}
%\label{f_x_rho2}
%\end{figure} 

%\begin{figure}
% \begin{center}
%\includegraphics[angle=0,width=0.5\linewidth]{x_rho.eps}
%\end{center}
%\caption{Charge density close to the left electrode (located at $x=-1$) as a function of the position for $\gamma=1$, $\mu=1$ and $Q=4.25$: exact result (solid line) and mean field result (dashed line).}
%\label{f_x_rho}
%\end{figure} 

%\begin{figure}
% \begin{center}
%\includegraphics[angle=0,width=0.5\linewidth]{x_rho_layers.eps}
%\end{center}
%\caption{Exact result for the mean charge density close to the left electrode (located at $x=-1$) as a function of the position for $\gamma=1$, $\mu=1$ and $Q=0.5$, for different values of the fugacity.}
%\label{f_x_rho_layers}
%\end{figure} 

%\begin{figure}
% \begin{center}
%\includegraphics[angle=0,width=0.5\linewidth]{x_rho_layers2.eps}
%\end{center}
%\caption{Exact result for the mean charge density as a function of the position for $\gamma=1$, $\mu=1$ and $Q=0.5$, for different values of the fugacity.}
%\label{f_x_rho_layers2}
%\end{figure} 

\section{Numerical Simulations}

Because of the intricacies observed in the behavior of this system we deem it advisable to check the theoretical results against simulation. In this section we present a selection of simulation results which cover the different aspects of the 1D LCF. The simulation is performed directly with the spin formulation and so the comparison serves also as a check of the Hubbard-Stratonovich transformation to the formulation in terms of the field potential $\phi$. In all cases agreement between theory and experiment is absolutely excellent. 

We can consider both the canonical and grand-canonical ensembles but concentrate here on the
grand-canonical ensemble which is appropriate to the scalar field reformulation. We simulate both for
the constant $Q$ and constant $\Delta v$ ensembles which are related by the usual Legendre
transformation. From earlier, on a lattice with $M$ sites, we have
\begin{eqnarray}
\beta H_0&=&-\frac{\gamma}{4}\sum_{i,j=0}^{M-1} |i-j|S_iS_j \nonumber\\
\beta H &=& \beta H_0 + \gamma u Q + \frac{1}{2}\gamma Q^2(M+1),
\end{eqnarray} 
where $u = \sum_{i=0}^{M-1}iS_i$. For the fixed $Q$ grand ensemble with chemical potential $\mu$ 
the partition function is
\begin{equation}
\Xi_Q = \sum_{\{S_p\}} \exp\left(-\beta H -\log \mu\sum_i|S_i|^2\right),
\end{equation}
and for the fixed $\Delta v$ grand ensemble the partition function is
\begin{eqnarray}
\Xi_{\Delta v}&=&\int dQ\;\Xi_Q\;\exp(Q \Delta v) \nonumber \\
&=& \sum_{\{S_p\}} \exp\left(-\beta H_0 + \frac{1}{2} \frac{(\Delta v + u)^2}{\gamma (M+1)}\right),
\end{eqnarray}
where $\Delta v$ is the fixed potential difference between the plates in the 
appropriate units.
 
The update is done using the Metropolis algorithm. We select a pair of neighboring
spins $S_i, S_{i+1}$ at random. If $|S_i + S_{i+1}| = 2$ then no update is possible because
of charge conservation and we select a new pair. In the case $|S_i + S_{i+1}| < 2$ we suggest
an update of the pair which respects charge conservation but is otherwise chosen with
equal probability; this preserves detailed balance. We choose this probability to obtain
an efficient acceptance rate. The suggested update is accepted/rejected using the standard Metropolis
procedure. One lattice update consists of $M$ pair updates. We typically perform of order
$10^5-10^6$ lattice updates to establish equilibrium and then measure the operator under
consideration every subsequent lattice update and compute the average at the end of the run. 
We typical use $10^5-10^6$ measurement lattice updates. All calculations were performed on 
multiple processor machines, each processor of which was used to carry out an independent 
simulation of the kind described here. The grand average over the independent simulations
was taken and the error computed from the variance of the distribution of the individual
processor averages. We used both a 12 core intel desktop computer and also an NVIDIA Tesla
GPU with 512 cores. Results from both these computers agree very well indeed.  

\begin{figure*}[t!]\begin{center}
	\begin{minipage}[b]{0.485\textwidth}
	\begin{center}
		\includegraphics[width=\textwidth]{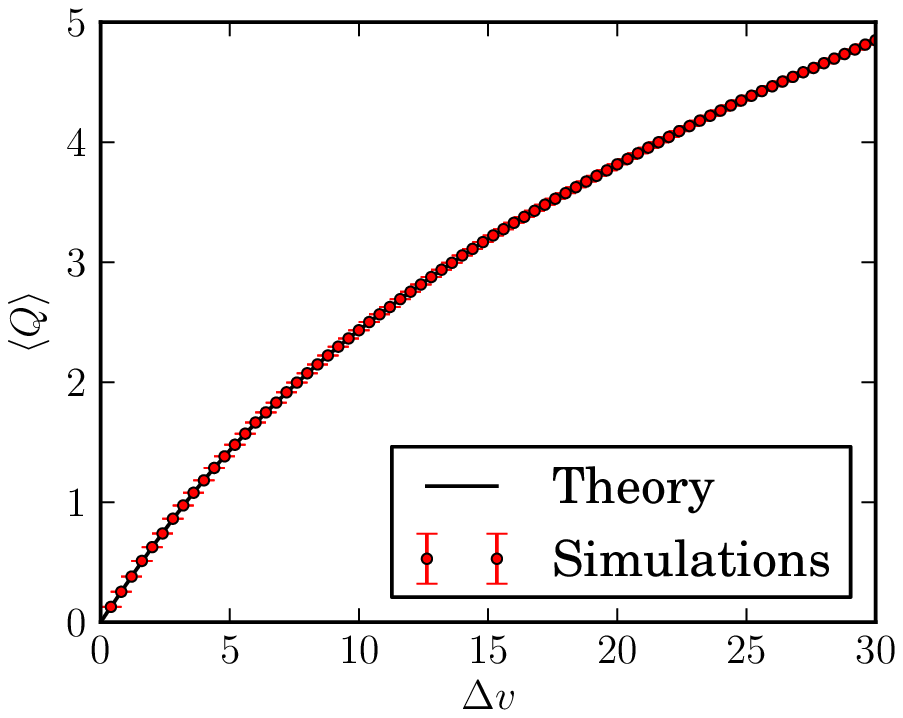}
	\end{center}\end{minipage} \hskip0.25cm
	\begin{minipage}[b]{0.485\textwidth}\begin{center}
		\includegraphics[width=\textwidth]{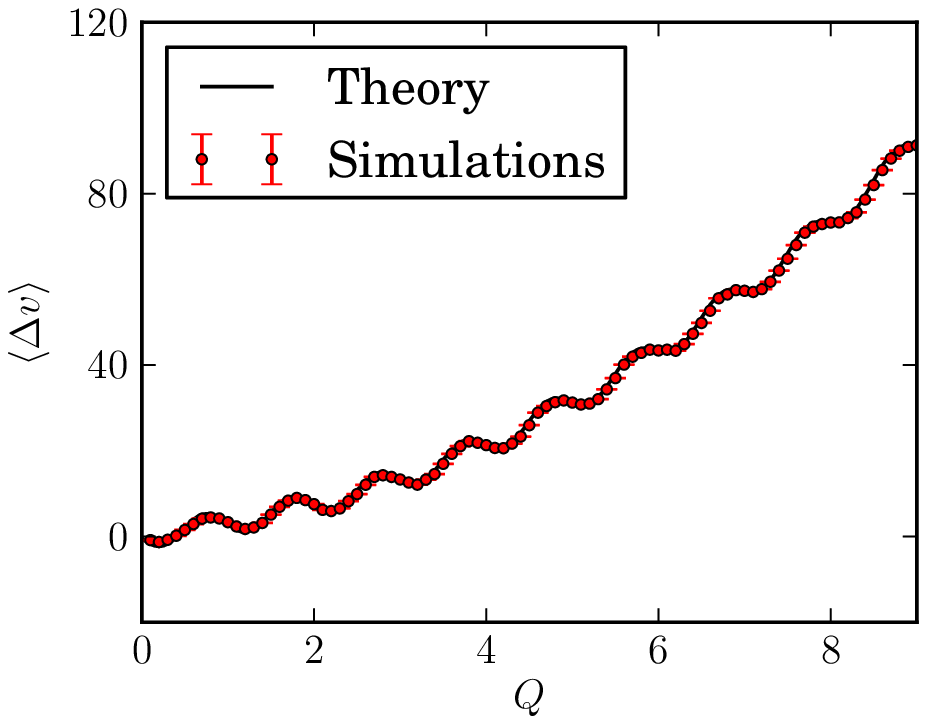}
	\end{center}\end{minipage} \hskip0.25cm	
\caption{Left: Theory and simulation for mean charge $\langle Q \rangle$ versus $\Delta v$ for
$\gamma=1, \mu=1$ on a lattice with $M=128$. Right: Theory and simulation for mean voltage difference $\langle \Delta v \rangle$ versus $Q$ for $\gamma=1, \mu=1$ on a lattice with $M=1024$. }
\label{V_Q_simu}
\end{center}\end{figure*}

In Fig \ref{V_Q_simu} we show $\langle Q \rangle$ versus $\Delta v$ in the $\Delta v$ 
ensemble for $\gamma=1, \mu=1, M=128$. We have
\begin{equation}
\langle Q \rangle = \frac{\partial}{\partial(\Delta v)} \log\;\Xi_{\Delta v} = 
    \frac{\Delta v + \gamma \langle u \rangle}{\gamma (M+1)}.
\end{equation}
The graph in the figure shows that theory and simulation agree very well. On the same figure
we also show $\langle \Delta v \rangle$ versus $Q$ in the $Q$ 
ensemble for $\gamma=1, \mu=1, M=1024$. Here we have
\begin{equation}
\langle \Delta v \rangle = -\frac{\partial}{\partial Q} \log\;\Xi_Q = 
    \gamma \left((M+1) Q - \langle u \rangle\right).
\end{equation}
Again the agreement of simulation with theory is extremely good.

\begin{figure}
 \begin{center}
\includegraphics[angle=0,width=0.5\linewidth, clip=true]{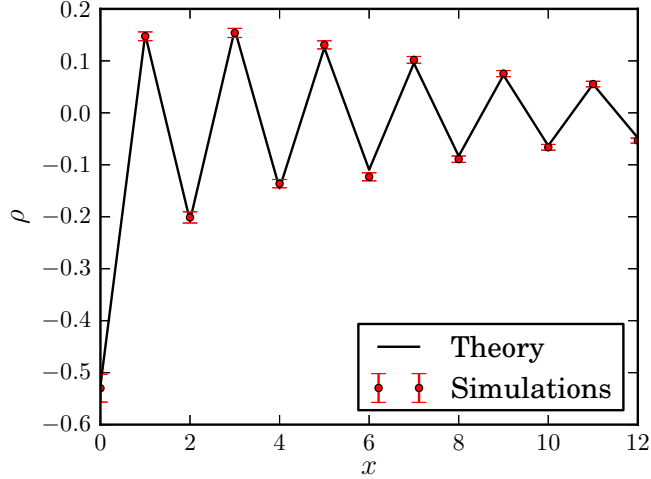}
\end{center}
\caption{Theory and simulation in the fixed $Q$ ensemble for the mean charge density $\rho$ 
versus distance $x$ from the left-hand plate located at $x=-1$ for $\gamma=1, \mu=10, Q=-0.5$ 
on a lattice with $M=1024$. Note that there is mild over-screening at $x=0$. }
\label{x_rho_simu}
\end{figure}

In Fig \ref{x_rho_simu} we show the mean charge density $\rho$ versus $x$ the distance 
in lattice units, $a$, from the left hand plate which carries charge $Q = 0.5$ and is 
located at $x=-1$. This is in the $Q$ ensemble for $\gamma=1, \mu=1, M=1024$. In this case
\begin{equation}
\rho = \langle S_x \rangle,~~ x = 0,1,2, \ldots\;.
\end{equation}
The layering of charge is clearly seen and the simulation faithfully reproduces the theory.
Mild over-screening at $x=0$ can also be seen where $\rho = -0.53$ which over-screens the charge of
$Q=0.5$ on the plate at $x=-1$.

\begin{figure*}[t!]\begin{center}
	\begin{minipage}[b]{0.485\textwidth}
	\begin{center}
		\includegraphics[width=\textwidth]{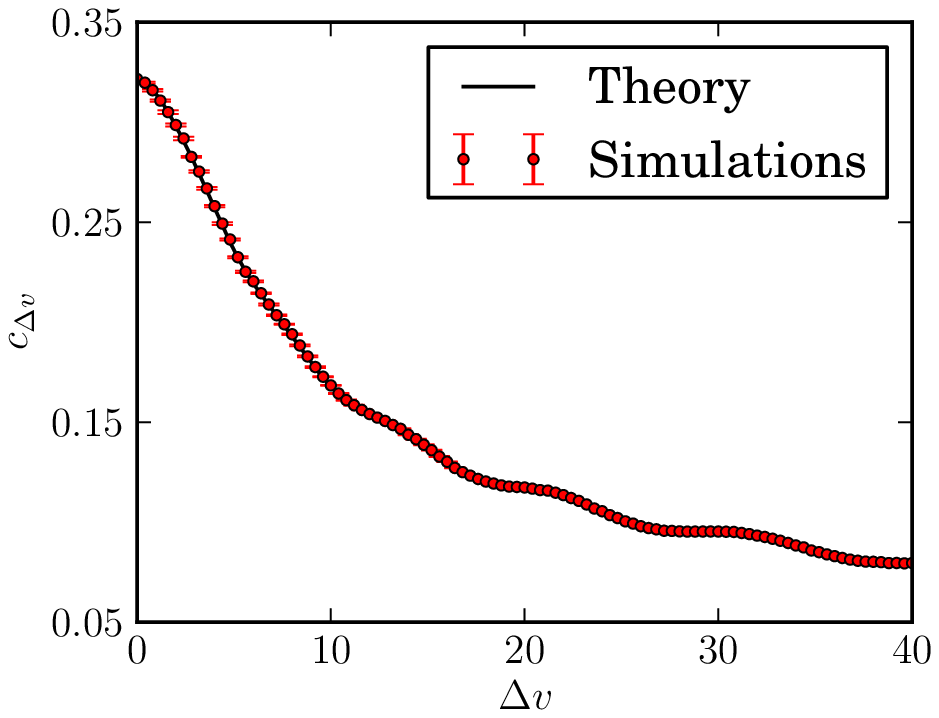}
	\end{center}\end{minipage} \hskip0.25cm
	\begin{minipage}[b]{0.485\textwidth}\begin{center}
		\includegraphics[width=\textwidth]{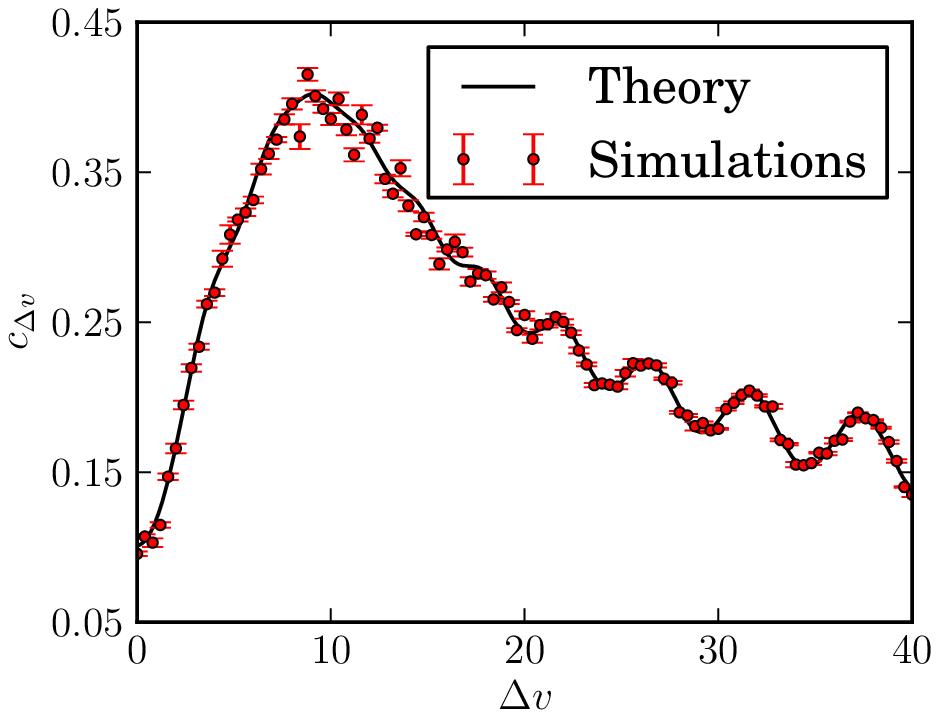}
	\end{center}\end{minipage} \hskip0.25cm	
\caption{Left: Theory and simulation for capacitance $C_{\Delta v}$ versus $\Delta v$ in the fixed $\Delta v$ 
ensemble for $\gamma=1, \mu=1$ on lattice with $M=128$. Right: Theory and simulation for capacitance $C_{\Delta v}$ versus $\Delta v$ in the fixed $\Delta v$ ensemble for $\gamma=0.3, \mu=0.03$ on lattice with $M=128$. }
\label{V_C_bell_simu}
\end{center}\end{figure*}

In Fig. \ref{V_C_bell_simu} we show the capacitance $C_{\Delta v}$
versus applied voltage $\Delta v$ in the fixed $\Delta v$ ensemble for $M=128$ and the respective 
cases $\gamma=1, \mu=1$ and $\gamma=0.3, \mu=0.03$. $C_{\Delta v}$ is time-consuming to calculate
since it is given in terms of the connected two point function $\langle u^2 \rangle_c$:
\begin{equation}
C_{\Delta v} = \frac{\partial^2}{\partial(\Delta v)^2} \log\;\Xi_{\Delta v} =
 \frac{1}{\gamma (M+1)} + \frac{\langle u^2 \rangle_c}{(M+1)^2}.
\end{equation}
Connected two point functions are generally hard to calculate since they are a small difference
between two large quantities for which we have separate statistical estimates. The cancelation gives 
an estimate that is a small number but with an error commensurate with that of each term separately. 
Our simulation results took well over 24 hours to complete which is why we were limited to the 
relatively small lattice size $M=128$. The results again agree well with the theory curve.

We conclude by stating that we can successfully simulate the model formulated in terms of lattice spins and
that the results agree very well with the formulation in terms of the potential field which is
given by the Hubbard-Stratonovich transformation. The agreement between theory and simulation
also verifies the Fourier method used in the exact theoretical analysis.

%\begin{figure}
% \begin{center}
%\includegraphics[angle=0,width=0.5\linewidth, clip=true]{V_Q_simu.eps}
%\end{center}
%\caption{Theory and simulation for mean charge $\langle Q \rangle$ versus $\Delta v$ for $\gamma=1, \mu=1$ on a lattice with $M=1024$.}
%\label{V_Q_simu}
%\end{figure}

%\begin{figure}
% \begin{center}
%\includegraphics[angle=0,width=0.5\linewidth, clip=true]{Q_V_simu.eps}
%\end{center}
%\caption{Theory and simulation for mean voltage difference $\langle \Delta v \rangle$ versus $Q$ for $\gamma=1, \mu=1$ on a lattice with $M=1024$.}
%\label{V_Q_simu}
%\end{figure}

%\begin{figure}
% \begin{center}
%\includegraphics[angle=0,width=0.5\linewidth, clip=true]{V_C_bell_simu.eps}
%\end{center}
%\caption{Theory and simulation for capacitance $C_{\Delta v}$ versus $\Delta v$ in the fixed $\Delta v$  ensemble for $\gamma=1, \mu=1$ on lattice with $M=128$.}
%\label{V_C_bell_simu}
%\end{figure}

%\begin{figure}
% \begin{center}
%\includegraphics[angle=0,width=0.5\linewidth, clip=true]{V_C_camel_simu.eps}
%\end{center}
%\caption{Theory and simulation for capacitance $C_{\Delta v}$ versus $\Delta v$ in the fixed $\Delta v$ b ensemble for $\gamma=0.3, \mu=0.03$ on lattice with $M=128$.}
%\label{V_C_camel_simu}
%\end{figure}

\section{Conclusion}

We have studied a 1D lattice Coulomb fluid model in a capacitor configuration. The model can be solved exactly, allowing us to compute all the relevant quantities, notably the
pressure, the capacitance and the charge distribution. Even though the system is
one dimensional and it would thus be difficult to make quantitive comparison with experimental systems,
we believe the results are pertinent to the field of ionic liquid capacitors for two main reasons. First
it allows a comparison of an exact result with the much used mean-field theory. It thus reveals which 
characteristics of the exact solution can be captured by the mean-field theory. It seems that while the mean field theory successfully predicts overall trends and orders of magnitude, it fails to pick up on oscillatory behavior in charge distributions and some rather exotic behavior in the charge-voltage capacitor
characteristics of the model. 

The study suggests that the simple LCF  model can lead to a variety
of phenomena which are somehow washed out by the mean field approach, charge distribution 
oscillations, over-screening and very exotic behavior of the capacitance. Because the behavior predicted 
by our exact solution was so unexpected we have verified our analytical predictions by comparing them 
with numerical simulations of the 1D model and have shown them to be accurate.   Another point which should be made is that in contrast to \cite{Kornyshev1,Borukhov} we developed a discrete mean-field theory on the lattice rather than taking a continuum  limit. This is an important point as we show that even the discrete mean-field theory does not reproduce the various oscillatory phenomena  mentioned above, which are thus not a simply consequence of a discrete lattice and the charge oscillations that we see are therefore due to {\sl correlations} rather than  being an artifact due to the introduction of a lattice.

In concluding, we should also emphasize that the 1D model has certain pathologies that are
not present in a three dimensional system, notably the lack of screening of the surface charges
in systems with large plate separations. If we would like to make a link with experimental systems it is clear that the simple picture of charged sheets would need to be modified. The state space of the charge on the sheets here is rather cartoon-like with just three values and just two bare weights for the presence of the charges $\pm 1$ or $0$ (given by the fugacity $\mu$). An extension of this model would be to
introduce a larger set of dynamical values for the charges on the sheets and it may even be possible to
find an effective one dimensional lattice model from a complete three dimensional model by exploiting the 
layering parallel to the plates that should be induced by an applied voltage, perhaps via renormalization 
group type arguments.

We finally note that the results of numerical simulations agree with those of the theoretical calculations and
therefore confirm the equivalence of the spin and scalar field formulations as well as the use of the Fourier method to carry out the exact theoretical calculation.

\section{Acknowledgement}

R.P. acknowledges support from ARRS through the program P1-0055 and the research project J1-0908 as well as the University of Toulouse for a one month position of Professeur invit ́e.


\begin{thebibliography}{99}
\bibitem{poded}P. Kekicheff, C. Holm and R. Podgornik (Eds.). Electrostatic effects in soft matter and biophysics, Kluwer Academic, Dordrecht (2001).
\bibitem{pooned}W. C. K. Poon and D. Andelman (Eds.). Soft condensed matter physics in molecular and cell biology.  Taylor and Francis, New York, London (2006).
\bibitem{review} A. Naji {\sl et al.}. Exotic Electrostatics: Unusual Features of Electrostatic Interactions between Macroions, in W.-B. Hu \& A.-C. Shi editors, Series in Soft Condensed Matter, Vol. 3, World Scientific, Singapore (2010). 
\bibitem{bo2005} H. Boroudjerdi et al. Phys. Rep. {\bf 416}, 129  (2005).
\bibitem{RP-SP} R. Podgornik and B. Zeks, J. Chem. Soc., Faraday Trans. 2 {\bf 84}, 611 (1988).
\bibitem{mo2005} A.G. Moreira A. Naji, S. Jungblut and R.R. Netz. Physica A {\bf  352}, 131 (2005).
\bibitem{review1} P.	Simon and Y. Gogotsi, Nat. Mater., {\bf 7},845 (2008).
\bibitem{ste95} G. Stell, J. Stat. Phys. {\bf 78}, 197 (1995).
\bibitem{yeh96} S. Yeh, Z. Zhou and G Stell, J. Chem. Phys. {\bf 100}, 1415, (1996).
\bibitem{jia02} J. Jiang, L. Blum, O. Bernard, J. Pausnitz and S.I. Sandler, J. Chem. Phys. {\bf 116}, 7977 (2002).
\bibitem{Andelman-rev} D. Ben-Yaakov, D. Andelman, R. Podgornik, D. Harries, J. Coll. Interf. Sci (2011).
\bibitem{art03} M.N. Artyomov, V. Kobolev and A.B. Kolomeisky, J. Chem. Phys {\bf 118}, 6394 (2003).
\bibitem{kob02}V. Kobolev, A.B. Kolomeisky and M.E. Fisher, J. Chem. Phys. {\bf 116}, 7589 (2002).
\bibitem{bro02} A. Brognara, A. Parda and L. Reatto, Phys. Rev. E {\bf 65}, 066113 (2002).
\bibitem{cia02} Ciach abd G. Stell, J. Chem. Phys. {\bf 117}, 8879 (2002).
\bibitem{kob202} V. Koboloev and  A.B. Kolomeisky, J. Chem. Phys. {\bf 117}, 8879 (2002).
\bibitem{Kornyshev1} A. A. Kornyshev, J. Phys. Chem. B {\bf 111}, 5545 (2007). 

\bibitem{Bazant-rev} M.Z. Bazant {\sl et al.},  Adv. Coll. Interf. Sci. {\bf 152}, 48 (2009). 

\bibitem{Kornyshev2} M. V. Fedorov and A. A. Kornyshev, J. Phys. Chem. B Letts. {\bf 112}, 11868 (2008). 
\bibitem{lot10} M.S. Loth, B. Skinner and B.I. Shlovskii, Phys. Rev. E {\bf 82}, 016107 (2010).
\bibitem{ski10}B. Skinner, M.S. Loth and B.I. Shklovskii, Phys. Rev. Lett. {\bf 104}, 128302 (2010).
\bibitem{Mezger} M. Mezger {\sl et al.}, Science {\bf 322} 424  (2008).
\bibitem{SFA} S. Perkin, T. Albrecht and J. Klein, Phys. Chem. Chem. Phys., {\bf 12}, 1243 (2010); K. Ueno {\sl et al.}, Phys. Chem. Chem. Phys., {\bf 12}, 4066 (2010); I. Bou-Malham and L. Bureau, Soft Matter, {\bf 6}, 4062 (2010).
\bibitem{AFM} R. Hayes, S. Z. El Abedin and R. Atkin, J. Phys. Chem. B,  {\bf 113}, 7049 (2009).
\bibitem{Perkin} S. Perkin {\sl et al.}, Chem. Commun. {\bf 47} 6572 (2011).


\bibitem{Bazant}  M. Z. Bazant, B. D. Storey, and A. A. Kornyshev, Phys. Rev. Lett. {\bf 106}, 046102 (2011).
\bibitem{Borukhov} I. Borukhov, D. Andelman, and H. Orland, Electrochem. Acta, {\bf 46}, 221 (2000).
\bibitem{Santangelo} C.D. Santangelo, Phys. Rev. E {\bf 73}, 041512 (2006).
\bibitem{Kanduc} M. Kandu\v{c} {\sl et al.}, J. Phys.: Condens. Matter {\bf 21} 424103 (2009).
\bibitem{v1} V. D\'emery, D.S. Dean, T.C. Hammant, R.R. Horgan and R. Podgornik, Europhys. Lett. 97, 28004 (2012).
\bibitem{edlen}{A. Lenard, J. Math. Phys. {\bf 26}, 82 (1961);  S.F. Edwards and A. Lenard, J. Math. Phys.
{\bf 3}, 778 (1962)}
\bibitem{aiz} M. Aizenman and J. Fr\" ohlich, J. Stat. Phys. {\bf 26}, 2 (1981).
\bibitem{dd1d} D.S. Dean, R.R. Horgan and D. Sentenac, {\bf 90} 899 (1998).
\bibitem{dd1d2} D.S. Dean, R.R. Horgan, A. Naji and R. Podgornik, J. Chem. Phys. {\bf 130}, 094504 (2009).
\bibitem{kiyo2007} K. Kiyohara and K. Asaka, J. Chem. Phys. {\bf 126}, 214704 (2007).
%\bibitem{Ben-Y} I. Borukhov, D. Andelman and H. Orland, Phys. Rev. Lett. {\bf 79}, 435 (1997). 
\bibitem{villain} J. Villain, J. Phys. (Paris) {\bf 36}, 581 (1975).
\bibitem{bask}  G. Baskaran and N. Gupte, Phys. Rev. B {\bf 30}, 432 (1984).
\bibitem{helm}H. L. F. von Helmholtz, Ann. Phys. {\bf 165},353, (1853).


\end{thebibliography}
\end{document}